\documentclass[lineno,pdflatex,sn-mathphys]{sn-jnl}% Math and Physical Sciences Reference Style
\jyear{2023}%

\begin{document}

\title[Large-Signal Behavior of Ferroelectric Micro-Electromechanical Transducers]{Large-Signal Behavior of Ferroelectric Micro-Electromechanical Transducers}

\author[1]{\fnm{Udit} \sur{Rawat}}

\author[2]{\fnm{Jackson} \sur{Anderson}}

\author*[2]{\fnm{Dana} \sur{Weinstein}}\email{danaw@purdue.edu}

\affil[1]{\orgdiv{Kilby Labs}, \orgname{Texas Instruments}, \orgaddress{\city{Dallas}, \state{TX}, \country{USA}}}
\affil[2]{\orgdiv{Department of Electrical and Computer Engineering}, \orgname{Purdue University}, \orgaddress{\city{West Lafayette}, \state{IN}, \country{USA}}}

\abstract{CMOS-MEMS resonators seamlessly integrated in advanced integrated circuit (IC) technology have the unique capability to enable unprecedented integration of stable frequency references, acoustic spectral processors, and physical sensors. Demonstrations of transducers leveraging piezoelectric properties of emerging ferroelectric materials such as Hafnium Zirconium Oxide (HZO) and Aluminum Scandium Nitride (AlScN) enable high figure of merit (FOM $=k^2Q$) resonators over a wide range of frequencies. CMOS-integrated ferroelectric transducers using a thickness-scaled variant of these films for low voltage operation are feasible by leveraging advancements in ferroelectric random access memory (FRAM) and ferroelectric field effect transistors (FeFETs). However, until now, there has not been a full treatment of the effects of nonlinear large-signal behaviour on the performance of electromechanical systems built using ferroelectric transducers in the low coercive voltage regime. In this work, CMOS-MEMS resonators in a 130 nm process operating at $\sim$700 MHz have been used as a vehicle for understanding the performance impact of nonlinear piezoelectric transduction on frequency references and acoustic filters. The first nonlinear large signal model for such resonators has been developed and employed to extract the nonlinear characteristics over different biasing and applied power. Operating conditions and design guidelines have also been developed for these applications, which can be extended to all resonators of this class. The crystallized understanding of large-signal operation of ferroelectric transducers presented in this work provides opportunities to design and demonstrate new capabilities of electromechanical devices in monolithic CMOS-MEMS platforms.}

%\keywords{keyword1, Keyword2, Keyword3, Keyword4}

\maketitle

\section{Introduction}\label{sec1:Intro}

Realization of monolithic MEMS components in advanced Complementary Metal Oxide Semiconductor (CMOS) IC manufacturing technology has the potential to enable new paradigms in next-generation communications, Internet of Things (IoT), and hardware security. Additional functionality pertaining to Radio Frequency (RF) systems and physical sensing with significant power, area, and cost advantages can be enabled in standard CMOS and emerging 3D heterogeneously integrated (3DHI) IC technology with minor or no modifications to manufacturing and packaging. The emergence of such monolithic solutions is timely in the face of enhanced pressure to integrate and shrink RF front-end module components \cite{Aigner20193G4} for current and future generation wireless communication (xG) standards. This challenge, caused by the introduction of new bands and filter-intensive techniques such as Carrier Aggregation (CA) and Multiple Input Multiple Output (MIMO) \cite{Fattinger2016CarrierAA}, is further exacerbated by a need for configurability. Current Surface Acoustic Wave (SAW) and Bulk Acoustic Wave (BAW) filters \cite{RFfilters1,RFfilters2,RFfilters3,RFfilters4}, due to their lack of configurability and requirement of specialized fabrication and packaging, are not suitable in this scenario. Barium Strontium Titanate (BST) Film Bulk Acoustic Resonators (FBARs) \cite{ZolfagharlooKoohi2020ReconfigurableRE,BST_FBAR} provide configurable filters due to their ferroelectric properties, but are not CMOS compatible. Intrinsically-switchable ferroelectric-transduced acoustic wave devices in commercial CMOS processes are well poised to address these problems through the elimination of RF switches from font-end modules and monolithic integration. Additionally, autonomous adaptive radio front ends capable of re-configuring their performance characteristics based on the level of the incoming signal can potentially be realized using ferroelectric transduction based MEMS resonators. Ferroelectric devices are currently also a subject of active research for use in neuromorphic compute paradigms to overcome the von Neumann bottleneck \cite{Ni2020FerroelectricsFM, Liu2022ReconfigurableCO, Dutta2020Monolithic3I, Han2022FerroelectricDF}. These ferroelectric-based solutions consist of hardware based on Spiking Neural Networks (SNNs) \cite{Peng2019NanocrystalEmbeddedInsulatorF} or Convolutional Neural Networks (CNNs) \cite{Kim2022CMOScompatibleCA} that leverage the innate capability of the underlying devices to perform storage and processing simultaneously resulting in implementation possibilities such as in-sensor computation \cite{Wan2022InSensorCM, Datta2022InSensorN, Zhu2021RecentAI}. Moreover, on-chip high frequency ultrasound generation through the use of CMOS-integrated BEOL ferroelectric transducers enables applications such as fingerprint detection \cite{USfingerprint}, high-resolution clinical imaging \cite{Fei2016UltrahighF}, laser-free picosecond ultrasonics for thin film metrology \cite{Matsuda2015FundamentalsOP}, and neuromodulation \cite{Balasubramanian2020GHzUC}, most of which have so far never been implemented in scalable IC manufacturing technologies. All of the aforementioned applications require the application of large signal drive to the transducer elements. Therefore, for the appropriate design of actuators and physical sensors using this technology an understanding of the large signal behavior of integrated ferroelectric transducers is required.

Switchable MEMS resonators based on ferroelectric transduction are currently the subjects of active research \cite{tharpe_173_2022,Ghatge2019AnUI,ghatge_nano-mechanical_2018,ghatge_non-reciprocal_2019,ghatge_30-nm_2020,tharpe_-plane_2021,hakim_ferroelectric--si_2021,AlScN_Rassay,Wang2020AFB,wang_high-temperature_2022} spurred by the demonstration of ferroelectricity in back-end-of-line (BEOL)/front-end-of-line (FEOL) compatible materials such as doped Hafnium Dioxide (HfO\textsubscript{2})\cite{boescke_ferroelectricity_2011,polakowski_ferroelectricity_2015,HfO2_Mller2015FerroelectricHO,HfO2_Park2021BinaryFO,HfO2_Schroeder2022TheFA} and AlScN \cite{fichtner_ferroelectricity_2020,fichtner_alscn_2019}. However, with unique packaging and fabrication considerations, such resonant devices are yet to be monolithically integrated into a CMOS platform. Unreleased resonators incorporated in commercial CMOS with no post-processing have previously been demonstrated \cite{Radhika_UnreleasedRes32nm,bahr_monolithically_2016,bahr_32ghz_2018,bahr_theory_2015,he_tunable_2020, he_ferroelectric_2018, he_switchable_2019}. Out of these, one class \cite{he_ferroelectric_2018,he_switchable_2019,he_tunable_2020} makes use of switchable piezoelectric transduction with Lead Zirconate Titanate (PZT) ferroelectric capacitors (FeCAPs) integrated in the BEOL of the HPE035 130 nm CMOS FRAM process from Texas Instruments \cite{HPE035_McAdams2004A6E,HPE035_Udayakumar2005IntegrationAB}. These high-Q FeCAP-based devices are prime candidates for monolithically integrated intrinsically switchable RF filters and low phase-noise frequency references.

Acoustic wave filters for the transmit path (TX) in RF front-ends encounter high power signals from the RF power amplifier (PA). Moreover, filter nonlinearity can give rise to the generation of harmonics and intermodulation products affecting the performance of the receiver chain (RX). In frequency references, nonlinearity in the frequency filtering resonant element degrades the phase noise. Additionally, for transducers in physical sensing, large displacement/momentum generation is required for high sensitivity which is typically accomplished through large signal drive. Since CMOS integrated ferroelectric resonators are being considered for these applications, analysing the effect of large signal swings and nonlinearity on the analog performance of ferroelectric transducers originally purposed for digital FRAM is paramount. The FeCAP resonant devices discussed in this work have previously only been investigated under a very limited range of operating conditions. In this article, we report on the large signal and temperature-dependent characteristics of CMOS-MEMS ferroelectric resonators. The analysis and conclusions presented have direct implications for acoustic spectral processing, oscillatory systems and physical sensors making use of ferroelectric transducers using current and upcoming materials (such as HZO) for CMOS FRAM \cite{FRAM_Mikolajick2020ThePT,FRAM_BEOL} and FeFET applications \cite{Khan2020TheFO, FeFET1_8268425, FeFET2}. This work is not limited to transducers integrated in System-in-Package (SiP), but also informs the design and analysis of standalone ferroelectric transducers.  

\section{Device Design and Acoustic Dispersion}\label{sec2:DevDesign}

\begin{figure}[t!]
\centering
\includegraphics[width=\textwidth]{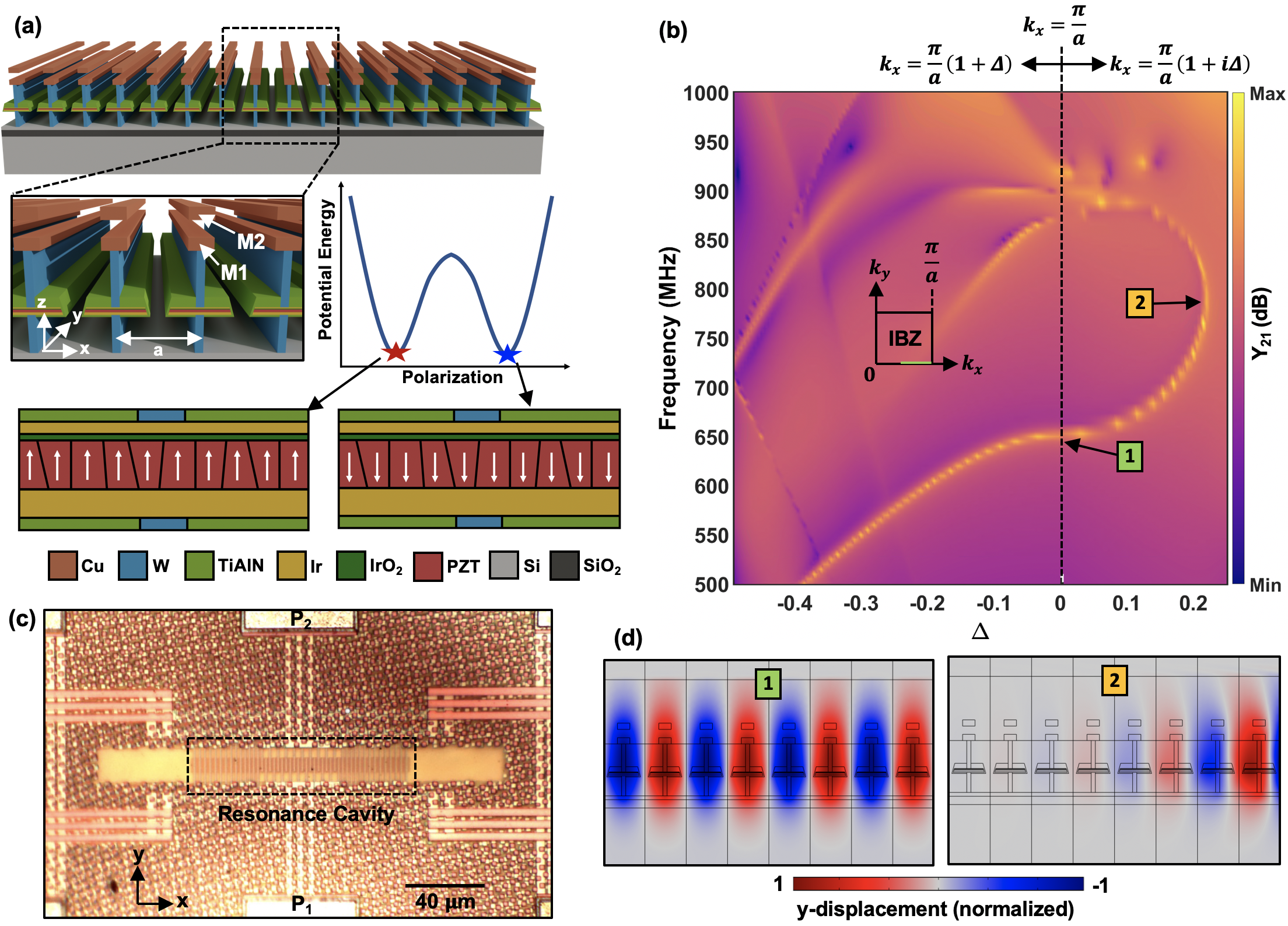}
\caption{(a) Schematic showing the phononic waveguide constructed using an array of FeCAP transducers. The individual unit cells are separated by a lattice constant $a$, which is lithographically defined and plays a dominant role in designing resonance frequency. The BEOL dielectric filling the space between the metals and FeCAPs has been omitted for visual clarity. The double well potential diagram (potential energy $U$ vs. polarization $P$) depicts two saturation states in the ferroelectric capacitor with opposite polarizations as shown. (b) Acoustic dispersion characteristics of Y\textsubscript{21} admittance for the waveguide evaluated using 2D FEM simulation. Regions to the left and right of the dashed line represent propagation and stopband characteristics corresponding to the electrically active modes supported by the waveguide, respectively. (c) Optical micrograph of the FeCAP resonator showing electrical connections to the transducers in the resonance cavity and the additional unit cells for reducing scattering on either side of the main cavity. (d) y-displacement mode shapes corresponding to target mode 1 resonance frequency and evanescent mode 2 in the stopband of the waveguide.}
\label{Fig1}
\end{figure}

The resonator fabricated in the HPE035 130 nm CMOS process \cite{HPE035_TI_FeRAM} from Texas Instruments, consists of a phononic waveguide grating formed using an array of BEOL PZT FeCAPs as shown in Fig.\ref{Fig1}(a). Each unit cell of the waveguide consists of a 70 nm thick ferroelectric PZT capacitor contacted with Iridium (Ir) and conductive Iridium Oxide (IrO\textsubscript{2}) electrodes. Tungsten (W) vias are used to connect the FeCAP cell to BEOL Cu routing and FEOL Polysilicon (Si) shield. The repetition of these unit cells along the x-direction (lattice constant $a=2$ $\mu$m) leads to the creation of a slow wave structure in which acoustic waves have a slower velocity of propagation as compared to the BEOL dielectric and the Si substrate underneath. Fig.\ref{Fig1}(c) shows the fabricated 108 $\mu$m by 7 $\mu$m FeCAP waveguide resonator cavity and metal routing to the signal and ground probe pads. To obtain the dispersion characteristics and mode coupling behaviour of the resulting phononic waveguide in terms of the transadmittance $Y_{21}$, Floquet Periodic Boundary Conditions (PBCs) are utilized in frequency domain analysis with periodic electrical excitation. Detuning the wave-vector $k_{x}$ by an amount $\Delta=\delta k_{x}/k_{x0}$ (both real an imaginary) along the $\Gamma-X$ path of the Irreducible Brillouin Zone (IBZ) gives propagation and stopband characteristics shown in Fig.\ref{Fig1}(b). The wave vector $k_{x0}=\pi/a$ corresponds to a phase shift of 180\textsuperscript{o} across the unit cell.

From the dispersion characteristics it is evident that guided mode 1 is capable of propagating in the slow-wave region with a grating-confined mode shape as shown in Fig.\ref{Fig1}(d). Additionally, the $\Delta-$frequency space indicates that no modes exist which can be detected electrically and driven efficiently in the waveguide in the immediate vicinity of Mode 1. This is relevant from the perspective of non-degenerate mode-coupling under large signal excitation of the resonance cavity \citep{modecoupling1_8672780} since it could lead to parametric effects such as mechanical domain frequency comb generation in CMOS. Identification of modes from the dispersion analysis that can be electrically excited and detected also informs the design of the high-Q resonance cavity. The resonance cavity is formed by termination of the grating, which introduces reflections. To minimize scattering of acoustic waves, we implement waveguide termination rather than an abrupt transition, in which evanescent Mode 2 (Fig.\ref{Fig1}(d)) in the termination waveguides is chosen with lattice constant so as to match the resonance frequency of Mode 1 in the cavity. The number of BEOL metal layers in the waveguide unit cell is restricted to two (M1 and M2) and the lateral dimensions are optimized to maximize the stress localization in the PZT thereby increasing the transduction efficiency.  In the device discussed in this work, lateral acoustic confinement in the $x-$direction is obtained by adding electrically disconnected dummy FeCAP unit cells to either side of the waveguide cavity so that acoustic scattering to the bulk occurs away from the transducer region.  

\section{Large Signal Performance Analysis}\label{sec3:LSanalysis}

\begin{figure}[t!]
\centering
\includegraphics[width=\textwidth]{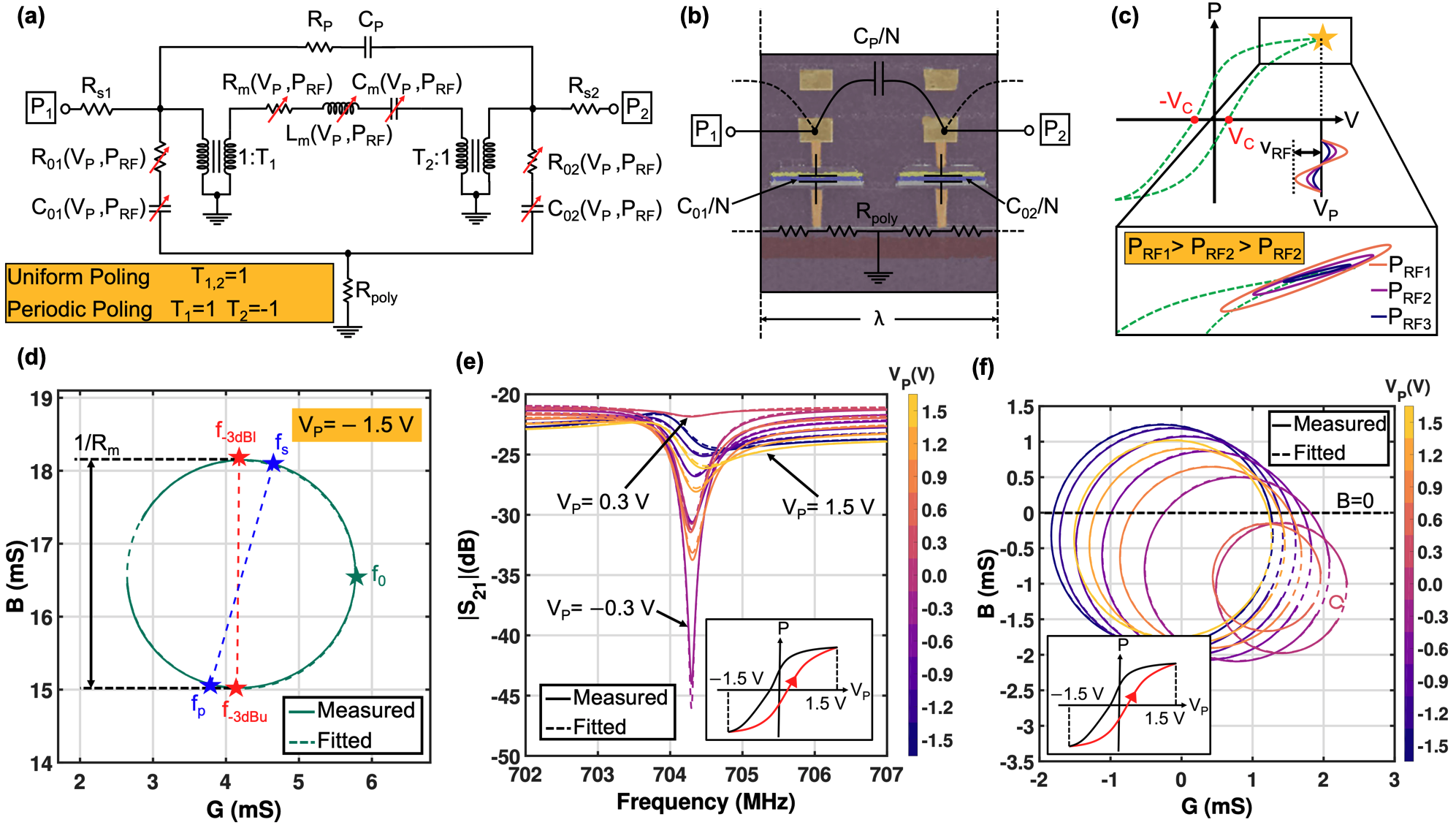}
\caption{(a) Two-port large signal equivalent circuit of the FeCAP resonator with $V_{P}$ and $P_{RF}$ dependent components. Transformers $T_{1}$ and $T_{2}$ are incorporated to ensure correct polarity of drive and sense transducers under uniform and periodic poling conditions \cite{he_switchable_2019}. (b) Cross-sectional Scanning Electron Micrograph of a pair of unit cells illustrating the origin of key electrical elements of the model. (c) Schematic representation of the effect of large signal drive on the $P-E$ loop hysteresis with applied bias voltage $V_{P}$. (d) Y\textsubscript{11} admittance circle for $V_{P}$= -1.5 V and $P_{RF}$= -10 dBm highlighting series and parallel resonance frequencies $f_s$ and $f_p$ corresponding to minimum and maximum admittance and -3 dB frequencies $f_{-3dBl}$ and $f_{-3dBu}$ associated with minimum and maximum susceptance $B$. (e) Bias-dependent $S_{21}$ transmission characteristics showing the close fit of the model to experiment. (f) $V_{P}$-dependent Y\textsubscript{21} admittance circles showing the biasing values for which the Barkhausen phase criterion is met with respect to $B=0$.}
\label{Fig2}
\end{figure}

Small-signal RF measurements across different biasing conditions as described in the Supplemental Information section \ref{supp_3} show significant variation in the device response with applied bias $V_{P}$. Conventional modified Butterworth-Van-Dyke (mBVD) modelling methods \cite{MBVD1_Larson2000ModifiedBD,MBVD2_Bjurstrom2004AnAD,MBVD3_Campanella2007AutomatedOC,MBVD4_Campanella2007InstantaneousDO,MBVD5_Erbes2014HighfrequencyPM} are not applicable for the FeCAP resonator performance extraction. This is primarily due to loss in modeling accuracy arising from insufficient circuit elements required to capture the observed characteristics. An augmented mBVD model as depicted in Fig.\ref{Fig2}(a) has therefore been used to accurately capture the resonator dependence on RF power ($P_{RF}$) and biasing ($V_{P}$). The motional branch representing Mode 1 is comprised of the motional resistance $R_{m}$, inductance $L_{m}$, and capacitance $C_{m}$, which are related through the mechanical resonance frequency $f_{0}$ (angular frequency $\omega_{0}$) and unloaded quality factor $Q_{UL}$. The physical origin of key parasitic elements in the resonator model is illustrated in Fig.\ref{Fig2}(b). FeCAP electrical loss elements $R_{01}$ and $R_{02}$ are introduced in series with the corresponding intrinsic transducer capacitances $C_{01}$ and $C_{02}$. These losses, just like the capacitances themselves, are dependent on bias and RF power. The series $RC$ combination of $R_{P}$ and $C_{P}$ captures the inter-finger capacitance and associated dielectric loss of the BEOL dielectric, and is considered invariant to bias and RF power. Meanwhile, $R_{poly}$ is included to model the resistance of the grounded polysilicon shield underneath the resonator structure connecting the bottom electrode of all ferroelectric transducers in the array.

Fig.\ref{Fig2}(c) clarifies the operating conditions under which the device is characterized. The resonator is biased at a polarization state (as indicated by the star in Fig.\ref{Fig2}(c)) using voltage $V_{P}$ superimposed with RF excitation voltage $v_{RF}$ corresponding to different $P_{RF}$. Under large-signal excitation, the polarization $P$ does not follow the characteristics typically seen when the FRAM is digitally switched between two states \cite{LSPZT2_Damjanovic1999CONTRIBUTIONSTT,LSPZT3_Taylor1998DomainWP,LSPZT4_Taylor1999NonlinearCT,LSPZT5_Gerber2004EffectsOF,LSPZT6_Zhang1988DomainWE,LSPZT7_Hall1998FieldAT}. Increasing $P_{RF}$ causes an increase in the average slope and area of the $P-E$ loop corresponding to each value. This can be attributed to loss arising from an increase in the irreversible domain wall motion as the large signal excursions are increased around the pinning voltage $V_{P}$. 

The measured $S_{21}$ transmission characteristics of the device at $P_{RF}=$-10 dBm and varying $V_{P}$ as seen in Fig.\ref{Fig2}(e) show resonance at approximately 704.1 MHz which corresponds to the mechanical resonance frequency $f_{0}$. The series resonance peak diminishes as $V_{P}$ moves from saturation towards the coercive voltage owing to an increase in the motional resistance $R_{m}$ (as evidenced by the reduction in admittance circle radii of Fig.\ref{Fig2})(f)). Due to change in transducer capacitances $C_{01}$ and $C_{02}$, there is a change in the relative magnitudes of electrical feedthrough current $i_{f}$ between the two ports and the motional current $i_{m}$ (brought about by corresponding variation in transducer electromechanical coupling). This results in variation in the resonator transfer function as detailed in \cite{he_switchable_2019,PZTres_Bedair2011HighRT} with a high-rejection parallel resonance obtained at $V_{P}$= -0.3 V corresponding to mutual cancellation of $i_{m}$ and $i_{f}$. In this particular device, the low amplitude of the series resonance is attributed to the uniform poling of alternate ferroelectric capacitors at the same bias voltage. As detailed in \cite{he_switchable_2019}, on the application of alternate poling voltages of opposing polarity to adjacent capacitors, the motional current dominates the feedthrough current resulting in a prominent series resonance. The fundamental mechanical Q-factor $Q_{UL}$ of the device is invariant to this poling condition. The parallel resonance under uniform poling shows maximum dependence on $V_{P}$ since transducer capacitance is strongly dependent on the applied bias which in turn modifies $i_{f}$. Admittance circle methodology (supplemental information section \ref{supp_2}) is used to extract augmented mBVD circuit parameters from measurement. Fig.\ref{Fig2}(d) shows the $Y_{11}$ Nyquist plot and the corresponding admittance circle fit to measured data for $P_{RF}=$-10 dBm and $V_{P}$= -1.5 V. Mechanical resonance frequency $f_{0}$, unloaded quality factor $Q_{UL}$ and the motional parameters are obtained for all $V_{P}$ and $P_{RF}$ combinations.  

\begin{figure}[t!]
\centering
\includegraphics[width=\textwidth]{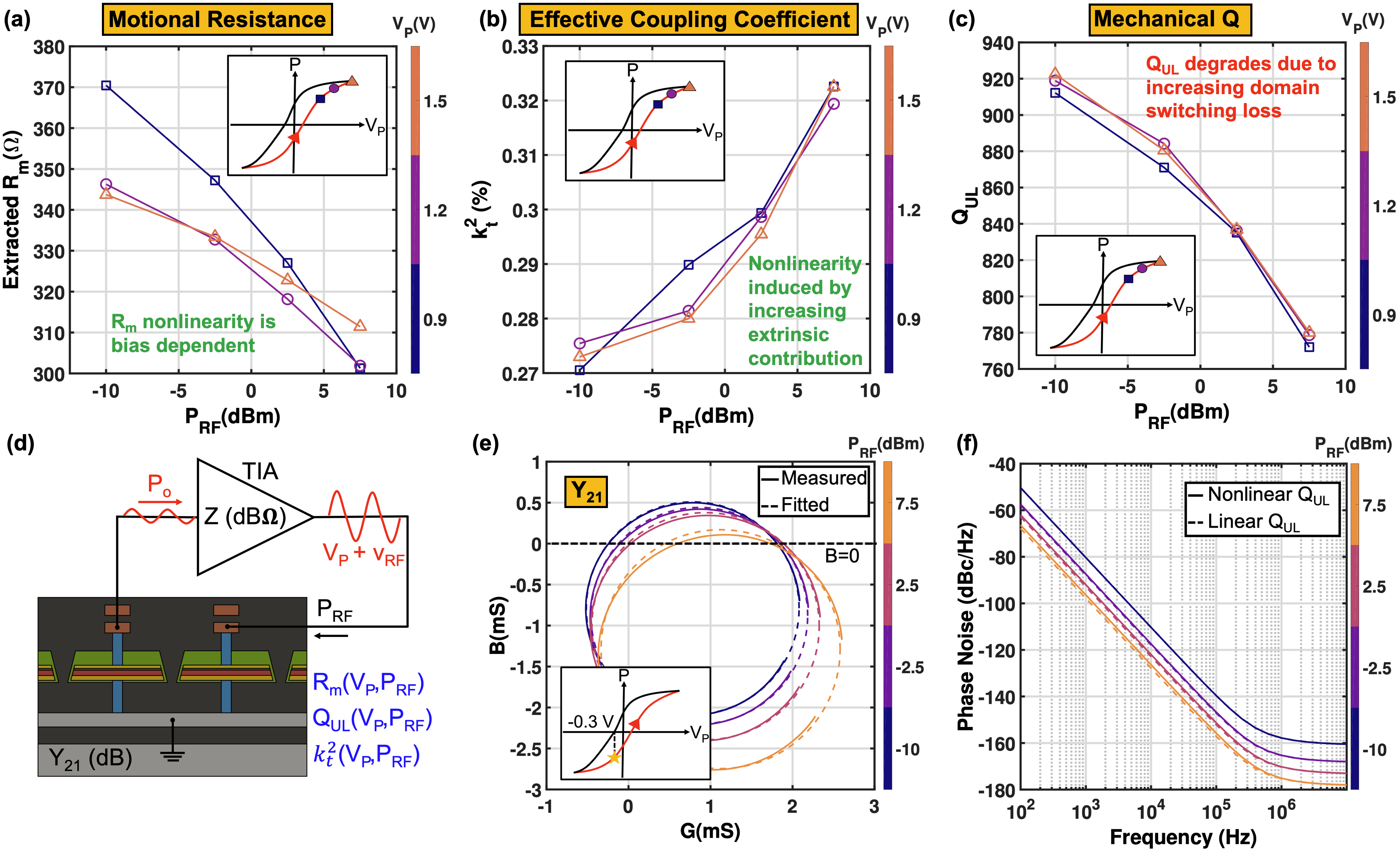}
\caption{$P_{RF}$-dependent nonlinearity of (a) motional resistance $R_{m}$, (b) mechanical quality factor $Q_{UL}$, and (c) effective electromechanical coupling coefficient ($k_{t}^2$). Isolation of nonlinear characteristics of these parameters enables understanding of the physical origins of large-signal behaviour. (d) Schematic of a TIA-based oscillator with gain $Z$. The FeCAP resonator admittance $Y_{21}(V_{P},P_{RF})$ is connected in feedback. (e) Dependence of the Y\textsubscript{21} admittance circle with $P_{RF}$ illustrating deviation from oscillation conditions for $V_{P}=-0.3$ V at higher power levels. (f) Simulated Harmonic Balance phase noise characteristics of the resonator biased at $V_{P}=$1.5 V with increasing $P_{RF}$ showing slight degradation in the phase noise due to $Q_{UL}$ nonlinearity.}
\label{Fig3}
\end{figure}

Illustrated in Fig.\ref{Fig2}(c), $P_{RF}$ levels of -10,-2.5,2.5 and 7.5 dBm are applied to the resonator at different biasing conditions ($V_{P}$) along the static $P-E$ loop. The resonator $R_{m}$, $Q_{UL}$, and effective coupling coefficient $k_{t}^2$ are then extracted using the augmented mBVD model for each of these operating points. From the extracted $R_{m}$ characteristics in Fig.\ref{Fig3}(a) for $V_{P}$ of 1.5, 1.2 and 0.9 V, we observe a reduction of 9.39\%, 12.92\%, and 18.65\% respectively for $P_{RF}$ increasing from -10 dBm to 7.5 dBm. Due to larger availability of switchable domains away from saturation, the extrinsic effects \cite{Li1991THEEN_extrinsic} in the FeCAP transducer increase resulting in stronger nonlinearity in $R_{m}$. Additionally, $Q_{UL}$ reduces with increasing $P_{RF}$ (Fig.\ref{Fig3}(c)) associated with an increase in irreversible domain wall motion caused by domain wall pinning \cite{revirrev2_Bolten1999ReversibleAI,revirrev1_Bolten2002ReversibleAI} at defect locations. This loss mechanism is prevalent in soft PZT used in FRAM. The increase in extrinsic ferroelastic effects also leads to an increase in the effective coupling coefficient of the resonator with $P_{RF}$ as seen in Fig.\ref{Fig3}(c). Non-180$^o$ domain walls are ferroelastically active and contribute 60-70$\%$ of the piezomodulii \cite{Li1991THEEN_extrinsic}; increasing the strength of the applied fields results in an increase in the motion of these domain walls, resulting in the cumulative increase in $k_{t}^2$.   

From the perspective of monolithic CMOS-MEMS oscillator design, the FeCAP resonator would generally be connected in the feedback path of a Transimpedance Amplifier (TIA) as shown in Fig.\ref{Fig3}(d) to form a closed loop. The $Y_{21}$ admittance circles (such as those in Fig.\ref{Fig3}(e) and (f)) determine whether the phase condition for oscillation is met. For sustained oscillations, the loop gain $Y_{21}Z$ should be greater than 1 and the $Y_{21}$ circle should intersect the susceptance $B$=0 line. Some admittance circles in Fig.\ref{Fig2}(f) lie completely below $B$=0, implying that it is not possible to obtain sustained oscillations at these bias voltages. For certain biasing points away from saturation such as $V_{P}$=-0.3 V in Fig.\ref{Fig3}(f), increasing $P_{RF}$ can cause the $Y_{21}$ circle to move below $B=0$ at which point oscillations are not possible. 

The oscillator phase noise is given by the Leeson's equation\cite{PhaseNoise_Lavasani2007A5L}:

\begin{equation}
    L(\Delta f) = \frac{FKT_{0}}{2P_{o}} \left[1+\frac{1}{f_{m}^2} \left(\frac{\Delta f}{2Q_{UL}}\right)^2 \right]\left(1+\frac{f_{\alpha}}{f_{m}}\right)
\end{equation}
where $f_{m}$ is the frequency of oscillation, $\Delta f$ is the frequency offset, $f_{\alpha}$ is a constant related to the $1/f$ noise corner, F is the noise figure of the TIA and $P_{0}$ is the oscillation power. For phase noise comparison, it is assumed that the TIA has zero input and output resistance, such that $Q_{UL} = Q_L$. Since the phase noise in the $1/f^{2}$ region depends on $Q_{UL}$, increasing $P_{0}$ reduces the phase noise at a slightly lower rate relative to the hypothetical case of a power-independent $Q_{UL}$. Fig.\ref{Fig3}(f) shows a comparison of simulated phase noise of an FeCAP CMOS-MEMS oscillator with and without power-dependent nonlinearity in the resonator. For $V_{P}$ = 1.5 V, power nonlinearity increases the phase noise at 1 kHz offset from carrier by 1.45 dB at 7.5 dBm RF power, assuming $Q$ is constant in both cases. Meanwhile, degradation in oscillator phase noise due to $Q_{UL}$ reduction with oscillation power is not substantial even for powers as high as 7.5 dBm, indicating viability of higher power operation limited instead by effects such as breakdown and fatigue in the ferroelectric material. The nonlinearity in the ferroelectric resonators gives rise to additional $1/f$ noise up-conversion near the carrier frequency requiring mitigation such as Automatic Gain Control (AGC) \cite{AGC_1269438}. Furthermore, due to an increase in $R_{m}$ close to the coercive voltage as seen in the Y\textsubscript{11} circle diameter in Fig.\ref{Fig2}(f), operating at these biasing conditions would require higher gain from the TIA and hence higher power consumption. 

\begin{figure}[t!]
\centering
\includegraphics[width=\textwidth]{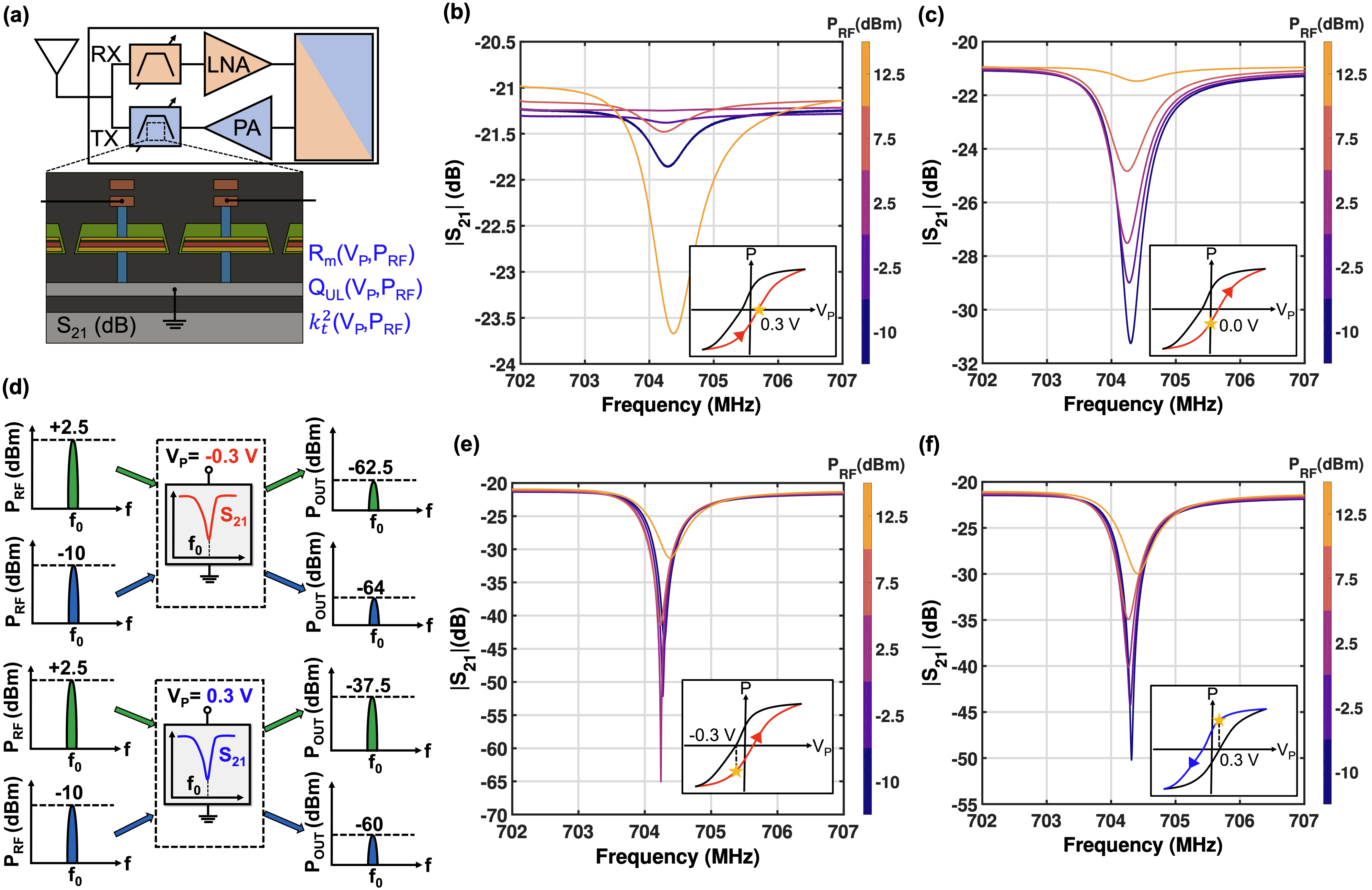}
\caption{(a) Schematic representation of an integrated RF transceiver module with switchable TX/RX duplexer using FeCAP resonators. Measured transmission characteristics when the resonator is biased at (b) the static coercive voltage $V_{C}$ = 0.3 V and (c) at $V_{P}$ = 0 V showing different switching characteristics dependent on applied $P_{RF}$. (d) Input signals at the parallel resonance frequency $f_{0}$ with different power levels (+2.5 and -10 dBm) undergo different levels of attenuation depending on the applied bias which needs to be dynamically adjusted for maximum rejection. Transmission characteristics at (e) $V_{P}$ = -0.3 and (f) 0.3 V on the $P-E$ loop as shown in the insets. As can be seen, the intensity of the parallel resonance dip in response caused by motional and feedthrough current cancellation is dependent on RF power.}
\label{Fig4}
\end{figure}

For RF front end applications, switchable multiplexers using FeCAP resonators can be monolithically integrated side by side with Low Noise Amplifiers (LNAs), Power Amplifiers (PAs), and other transmit and receive chain electronics for ultimate miniaturization (Fig.\ref{Fig4}(a)). However, the power handling and linearity of the FeCAP devices at the heart of the filters interfacing with the antenna and the amplifiers must be considered. From the perspective of switching capability, the resonators are expected to exhibit no electromechanical response when biased at the coercive voltage of the ferroelectric transducers. In this case, when the resonators are biased at the coercive voltage extracted from the static $P-E$ loop $V_{P}$=0.3 V, it is observed that the response nullification depends on $P_{RF}$ (Fig.\ref{Fig4}(b)). As RF power increases, the domain wall motion and switching increase causing a $V_{P}$-dependent nullification in the response. As an example, Fig.\ref{Fig4}(b) shows the response nullification for a $V_{P}$ of 0.3 V and $P_{RF}$ of 2.5 dBm. FeCAP filter switching is therefore dependent on the signal level; to ensure effective filter switching the bias would need to be adjusted according to the signal level. As an example, when the device is switched off for an input power of 12.5 dBm, $V_{P}$= 0 V must be applied as depicted in Fig.\ref{Fig4}(c). 

Modern concurrent wireless communication standards necessitate rejection of blocker signals with levels as high as 0 dBm, which can saturate receivers. Previous implementations of tunable bandstop or notch filters are area-intensive and primarily based on either low-loss substrate waveguides \cite{tunableNotch2, tunableNotch3} or integrated passive devices (IPDs) on Silicon \cite{tunableNotch1} coupled with RF MEMS elements. It has previously been suggested that the sharp parallel resonance notches common to many piezoelectrically and electrostatically transduced resonators can be utilized for blocker rejection in RF front ends \cite{PZTres_Bedair2011HighRT}. However, the FeCAP resonators in this work are uniquely capable of being used for blocker rejection in monolithic CMOS-MEMS RF front ends. The parallel resonance notch in the FeCAP resonators is most pronounced when biased at $V_{P}$ of 0.3 V or -0.3 V as shown in Fig.\ref{Fig4}(d) and (e) with rejection levels as high as 43 dB for the latter. From the transmission response of the resonator for these two biasing conditions in Fig.\ref{Fig4}(d) and (e), it can be seen that the notch depth varies with the input power level. Therefore, as the blocker signal level at the notch frequency changes from -10 to 2.5 dBm, the corresponding suppression (notch depth) changes from 28 to 18 dB with the FeCAP resonator biased at 0.3 V. Just as in the case of filter switching, adaptive biasing is required to reject the blocker based on its strength owing to the $P_{RF}$-dependent cancellation of motional and feedthrough currents.

For filters interfaced with the LNA and the PA, the insertion loss (IL) and out-of-band (OOB) rejection are governed by the Q of the resonators comprising the filters \cite{RFfilters2}. FeCAP resonator Q must therefore be maximized for IL and OOB rejection. However, from Fig.\ref{Fig3}(b) we see that resonator Q reduces with increasing $P_{RF}$. This means that a TX FeCAP resonator-based filter driven by the high $P_{RF}$ from the PA would experience degraded IL and OOB rejection as compared to the RX filter. This in turn requires an increase in current in the PA to compensate for the added loss, which is detrimental for system power consumption. Additionally, considering that transmitters in modern RF systems have very strict emission requirements, modifications to the filter passband edges due to $P_{RF}$ dependent nonlinearity of resonator $k_{t}^2$ (Fig.\ref{Fig3}(c)) places strict restrictions on the amount of power that can be delivered to the antenna. 

\begin{figure}[t!]
\centering
\includegraphics[width=\textwidth]{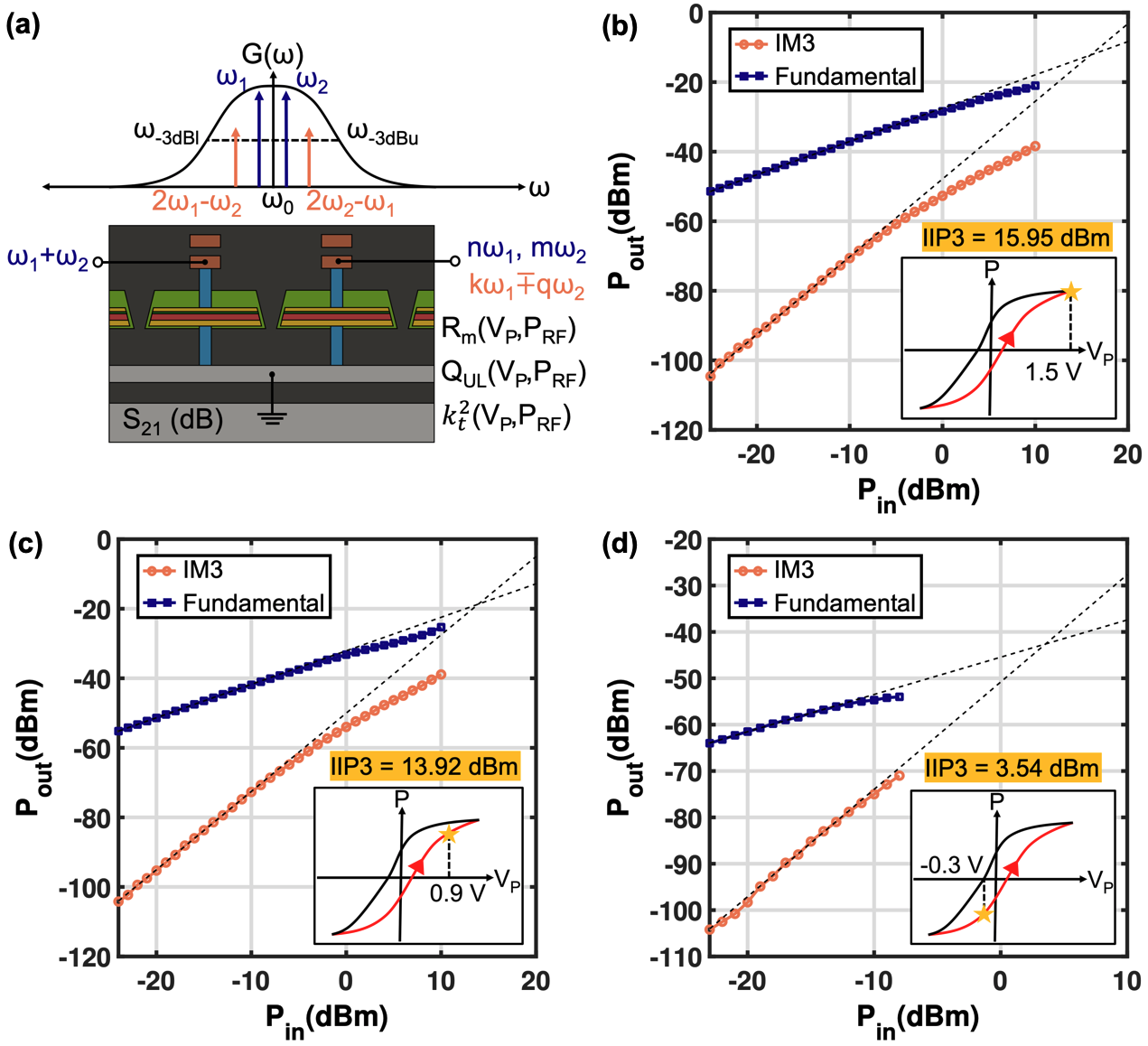}
\caption{ (a) Schematic representation of the two-tone ($\omega_{1}$ and $\omega_{2}$) test of the FeCAP resonator ensuring that the IM3 components lie within the half power bandwidth of the mechanical resonance, demarcated by frequencies $\omega_{-3dBl}$ and $\omega_{-3dBu}$. Measured IIP3 characteristics for biasing conditions corresponding to the (b) most linear (1.5 V), (c) bandstop (-0.3 V) and (d) intermediate (0.9 V) states. The progressively increasing nonlinearity is manifested as a reduction in the IIP3. The primary implication is that biasing the resonator away from saturation at 1.5 V for bandstop applications comes at the cost of increased harmonic and intermodulation component generation.}
\label{Fig5}
\end{figure}

Another aspect associated with the nonlinearity of the FeCAP resonators is the generation of harmonics and intermodulation products in the filters designed using these devices. Generation of such byproducts due to device nonlinearity is highly problematic for the receiver signal chain. To compare the nonlinearity at different biasing conditions, the resonator is characterized for the third order intercept point (IIP3) at three different $V_{P}$ values of 1.5 V, 0.9 V and -0.3 V. Two fundamental tones $\omega_{1}$ and $\omega_{2}$ spaced 50 kHz apart from the resonance frequency $\omega_{0}$ are applied to the biased resonator and the  third order intermodulation component (IM3) levels at $2\omega_{1}-\omega_{2}$ and $2\omega_{2}-\omega_{1}$ are recorded as shown in Fig.\ref{Fig5}(a). As can be seen from Fig.\ref{Fig5}(b), (c) and (d) the device exhibits IIP3 of 15.95 dBm, 13.92 dBm and 3.54 dBm for $V_{P}$ of 1.5, 0.9 and -0.3 V, respectively. The reduction in IIP3 with bias can be attributed to a shift away from the relatively linear saturation region where less domains are available for switching to the highly nonlinear region where the extrinsic ferroelectric characteristics are more pronounced. Thus, resonators biased in the off-state ($V_{P}$= 0 V) and bandstop ($V_{P}= -0.3 V$) configurations can exhibit harmonic and IM product generation which must be considered when designing filters using ferroelectric transducers. Device linearity can be improved significantly by utilizing resonator cascading techniques as demonstrated in \cite{linearityImprovement} opening opportunities for monolithic CMOS-MEMS RF front ends. On the other hand, the inherent nonlinearity and hysteresis of the ferroelectric electromechanical transducers can be exploited for autonomous adaptive radio front ends capable of tailoring their transfer functions depending on the input signal level. In RF sampling systems that make use of low dynamic range wide-band Analog-to-Digital Converters (ADCs), there is a need for cancellation of signals arising out of both internal spurs as well as external interferers. In all such applications the number of control inputs for the required re-configurable filters would be extremely large. Therefore, the signal-level dependent nonlinear behaviour of FeCAP resonators can be leveraged to alleviate this problem.

Aside from applications such as switchable acoustic filtering and frequency references, the nonlinearity, bias-dependent transduction, and hysteresis of ferroelectric materials can also enable new paradigms such as in-sensor computation in a conventional CMOS platform. Thus far, applications of in-sensor computing \cite{Chai2020InsensorCF} involving local information processing at the sensor-level for data efficiency have been limited to either photo-sensors \cite{Cui2021FerroelectricPN} or ferroelectric HZO-based released NEMS devices \cite{Jadhav2023ProgrammableFH}. Nonlinear ferroelectric MEMS/NEMS devices provide a significantly lower leakage pathway for in-memory/in-sensor computation is made feasible as compared to Fe-FET or RRAM based approaches \cite{Jadhav2023ProgrammableFH}. Through the use of nonlinear bias-dependent devices in this work this capability can be extended in CMOS to the acoustic domain for acoustic imaging and impedance spectroscopy. As an example, fingerprint sensing systems based on nonlinear ferroelectric transducer arrays implemented as a part of artificial neural networks can significantly increase efficiency by offloading computation from the digital processing circuits to the neuromorphic compute domain. Nonlinearity in the transduction also enables a plethora of applications associated with nonlinear parametric effects. Phononic frequency combs based on nonlinear ferroelectric transduction such as those demonstrated in \cite{Park2018PhononicFC,Park2019FormationEA} can enable unprecedented chip-scale spectrum sensing and metrology with simplified implementation as compared to optical combs. Additionally, parametric amplification exploiting ferroelectric nonlinearity can also be used to enhance the sensitivity of physical sensors as demonstrated using capacitive nonlinear transduction in a MEMS gyroscope \cite{Sharma2012ParametricRA,Hu2011APA}.

\section{Temperature Dependence}\label{sec4:tempVar}

\begin{figure}[t!]
\centering
\includegraphics[width=\textwidth]{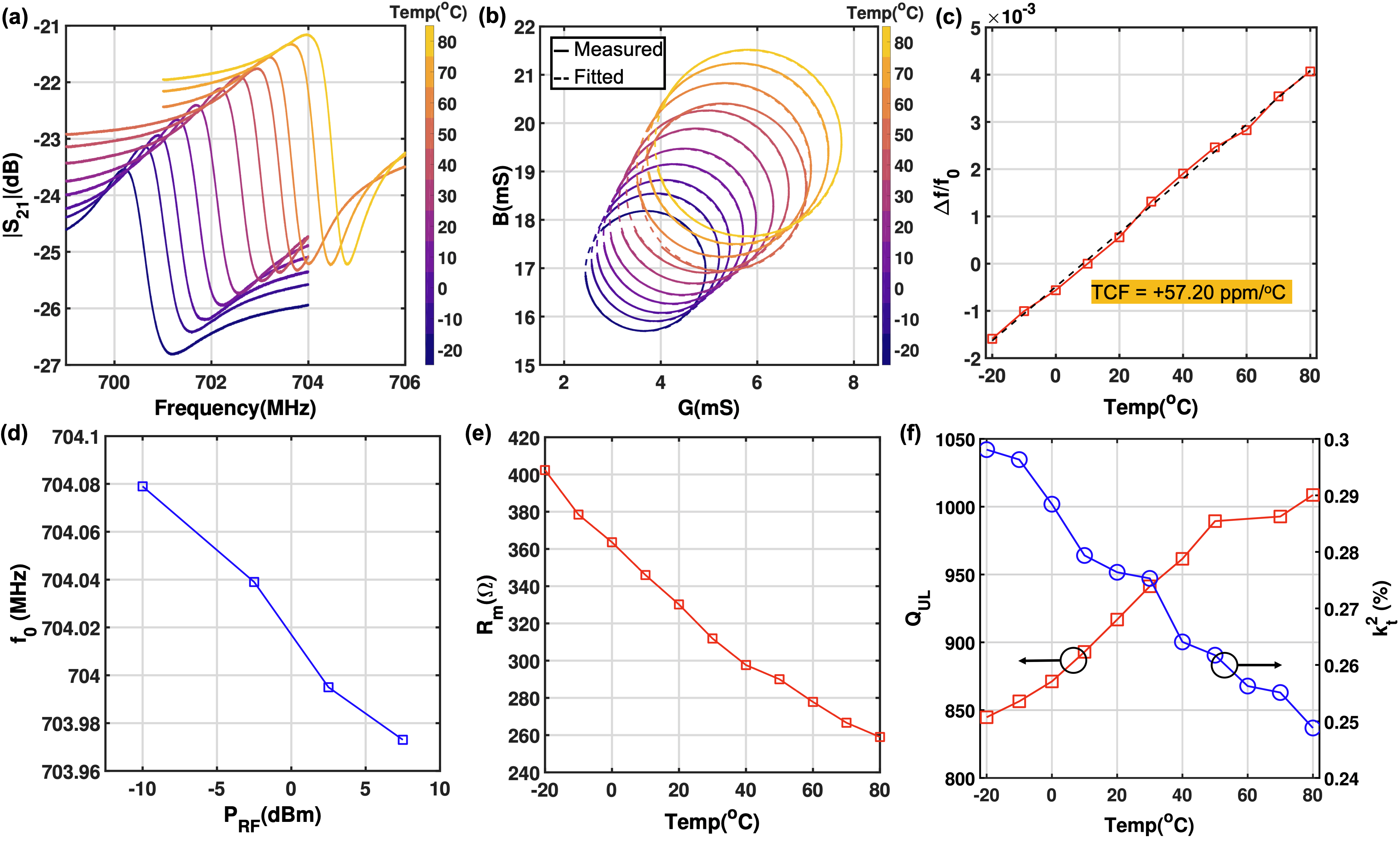}
\caption{(a) Temperature variation of the measured S\textsubscript{21} transmission characteristics from -20 to 80\textsuperscript{o}C depicting a progressive improvement in IL with increasing temperature. (b) Increase in the Y\textsubscript{11} admittance circle radius for $V_{P}$ = 1.5 V with increasing temperature. (c) Measured TCF of the FeCAP resonator for the resonance corresponding to 1.5 V bias. Variation of (d) Resonance frequency $f_{0}$ (e) Motional resistance $R_{m}$ (f) Effective electromechanical coupling coefficient $k_{t}^2$ and mechanical Q $Q_{UL}$ with temperature. Extracted temperature-dependent resonator parameters can be used to decipher the cause for large signal nonlinearities in the device.}
\label{Fig6}
\end{figure}

For temperature characterization, the resonators were tested under vacuum conditions in a Cascade PMC probe station with the temperature varied from -20 to 80 \textsuperscript{o}C. It should be noted that the device does not exhibit resonance characteristics around cryogenic temperatures of 77 K, consistent with what has been observed for other ferroelectric transducers \cite{cryoPZT1_Zhang1983DielectricAP,cryoPZT2_Gerson1962PiezoelectricAD,cryoPZT3_Thiercelin2010ElectromechanicalPO}. The measurements were performed under low RF power of -10 dBm to minimize self-heating due to domain wall motion. From Fig.\ref{Fig6}(a) and (b) obtained at a bias voltage of 1.5 V, we see that resonance frequency increases with temperature with a Temperature Coefficient of Frequency (TCF) of +57.20 ppm/\textsuperscript{o}C. Meanwhile, the IL reduces progressively with increasing temperature in the S\textsubscript{21} characteristics of Fig.\ref{Fig6}(a) corresponding to a reduction in the device motional resistance $R_{m}$. From the Y\textsubscript{11} admittance circles in Fig.\ref{Fig6}(b) we observe an increase in radius with temperature also in accordance with $R_{m}$ reduction. Additionally, the movement of the circles towards the top right points to an increase in the transducer capacitance $C_{0}$ as the temperature increases. Fig.\ref{Fig6}(e) shows the decrease in extracted $R_{m}$ and corresponding increase in piezoelectric coupling $k_{t}^2$ with increasing temperature.  

The mechanical Q $Q_{UL}$ exhibits an increase with temperature as shown in Fig.\ref{Fig6}(f). This behaviour is attributed to an increase in the mechanical Q of the ferroelectric PZT layer \cite{PZTQm1_Satoh1998TemperatureDO,PZTQm2_Miclea2007EffectOT} which bolsters $Q_{UL}$. From the obtained temperature and high RF power characteristics of the resonator obtained through extraction, a more complete picture of the resonator characteristics can be developed. First, the shift in resonance frequency due to increasing $P_{RF}$ is investigated. Possible origins of the shift include self heating caused by increasing RF amplitude or a change in stiffness. From Fig.\ref{Fig6}(d) we see that resonance frequency $f_{0}$ reduces as $P_{RF}$ increases, which is the opposite trend to the temperature-dependent increase seen in Fig.\ref{Fig6}(c). We conclude that the negative shift in the resonance frequency occurs due to a reduction in stiffness associated with the mode of interest. The physical origin of the reduction in motional resistance $R_{m}$ with increasing $P_{RF}$ can be traced to three mechanisms \cite{Wang2015PiezoelectricNI}. To better understand these mechanisms, let us consider $R_{m}$ in terms of the equivalent motional parameters of the resonance mode:

\begin{equation}\label{eq:RmNL}
    R_{m} = \frac{\sqrt{k_{eff}m_{eff}}}{Q_{UL}\eta_{1}\eta_{2}}
\end{equation}
where $k_{eff}$ and $m_{eff}$ are the effective stiffness and mass associated with the mode and $\eta_{1}$ and $\eta_{2}$ are the electromechanical coupling coefficients associated with the input and output transducers. Any reduction in $R_{m}$ is caused by a reduction in stiffness $k_{eff}$, increase in $Q$, and increase in electromechanical coupling coefficients $\eta_{1}$ and $\eta_2$. We have established a stiffness reduction with increasing $P_{RF}$. In addition, the PZT piezoelectric coefficients $d_{31}$ and $d_{33}$ were experimentally determined to exhibit RF amplitude dependence, and follow the Rayleigh equation \cite{LSPZT4_Taylor1999NonlinearCT}: 

\begin{equation}\label{eq:d33}
    d_{33} = d_{init} + \alpha E_{RF}
\end{equation}
where $d_{init}$ represents the reversible domain wall contribution to the electromechanical response, $\alpha$ is the Rayleigh constant and $E_{RF}$ is the amplitude of the electric field. Since the electromechanical coupling coefficients $\eta_{1}$ and $\eta_{2}$ are directly proportional to the piezoelectric coefficients, we can expect them to increase with $P_{RF}$. Therefore, the reduction in $R_{m}$ can be attributed to $\eta_{1}$ and $\eta_{2}$ alongside the reduction in stiffness with $P_{RF}$. Improvement in $R_{m}$ with increasing temperature can also be explained in a similar manner where there is an increase in the $Q_{UL}$. 

\section{Conclusion}\label{sec5:conclusion}

This work represents the first study of large signal characteristics in ferroelectric CMOS-MEMS resonators monolithically embedded in a commercial BEOL process stack. The primary application scenarios involve but are not limited to carrier generation and adaptive RF front-end filtering. Large signal treatment of ferroelectric transducers originally intended for low voltage operation in FRAM/FeFET is applicable to all such devices currently under development for the aforementioned applications. The first nonlinear large signal model and extraction methodology has been developed to obtain the performance characteristics under a wide range of biasing and RF power conditions. The inherent nonlinear mechanisms present in the electromechanical transduction can be leveraged for applications such as in-sensor computation and autonomous adaptive radio front-ends. It has been shown that the FeCAP resonator undergoes a reduction in both the $R_{m}$ and $Q_{UL}$ with $P_{RF}$ due to irreversible extrinsic electromechanical transduction unique to ferroelectrics. The performance of oscillators making use of these resonators is discussed in the context of phase noise and power dissipation, while considering the effect of nonlinearities. The influence of nonlinear behaviour on the design of switchable RF front end filters has also been illustrated through the use of IIP3 and $P_{RF}$-dependent characterization to set the optimal point for passband and bandstop type configurations. Finally, temperature dependence has been characterized tying together the understanding of performance variation under high $P_{RF}$. Analysis and design guidelines presented in this work for CMOS-MEMS ferroelectric resonators can facilitate additional functionality in standard RF CMOS and emerging 3D heterogeneously integrated (3DHI) ICs with minor or no modifications to manufacturing and packaging thereby enabling ultimate miniaturization. 

\section{Supplementary}\label{Supplementary}

\subsection{Modeling Background and Challenges}\label{supp_1}

To construct an appropriate strategy for the modeling of unreleased CMOS-MEMS ferroelectric resonators, the applicability of previously developed techniques for conventional SAW/BAW and ferroelectric devices must be evaluated. Three primary modeling approaches typically used for acoustic devices \cite{RFfilters1} include 1-D acoustic models such as Mason's model, P-Matrix and the Coupling-of-Modes (COM)-model based on constitutive equations, behavioural models like the modified Butterworth-van-Dyke (mBVD) model and black-box Poly Harmonic Distortion (PHD) model. For the unreleased resonant ferroelectric device under consideration, we first discuss appropriate model selection based on challenges of bias-dependent small-signal modeling and extraction. Subsequently, we address issues pertaining to the expansion of the chosen small-signal modeling methodology to incorporate the large-signal behaviour.

\begin{figure}[ht]
\centering
\includegraphics[width=0.8\textwidth]{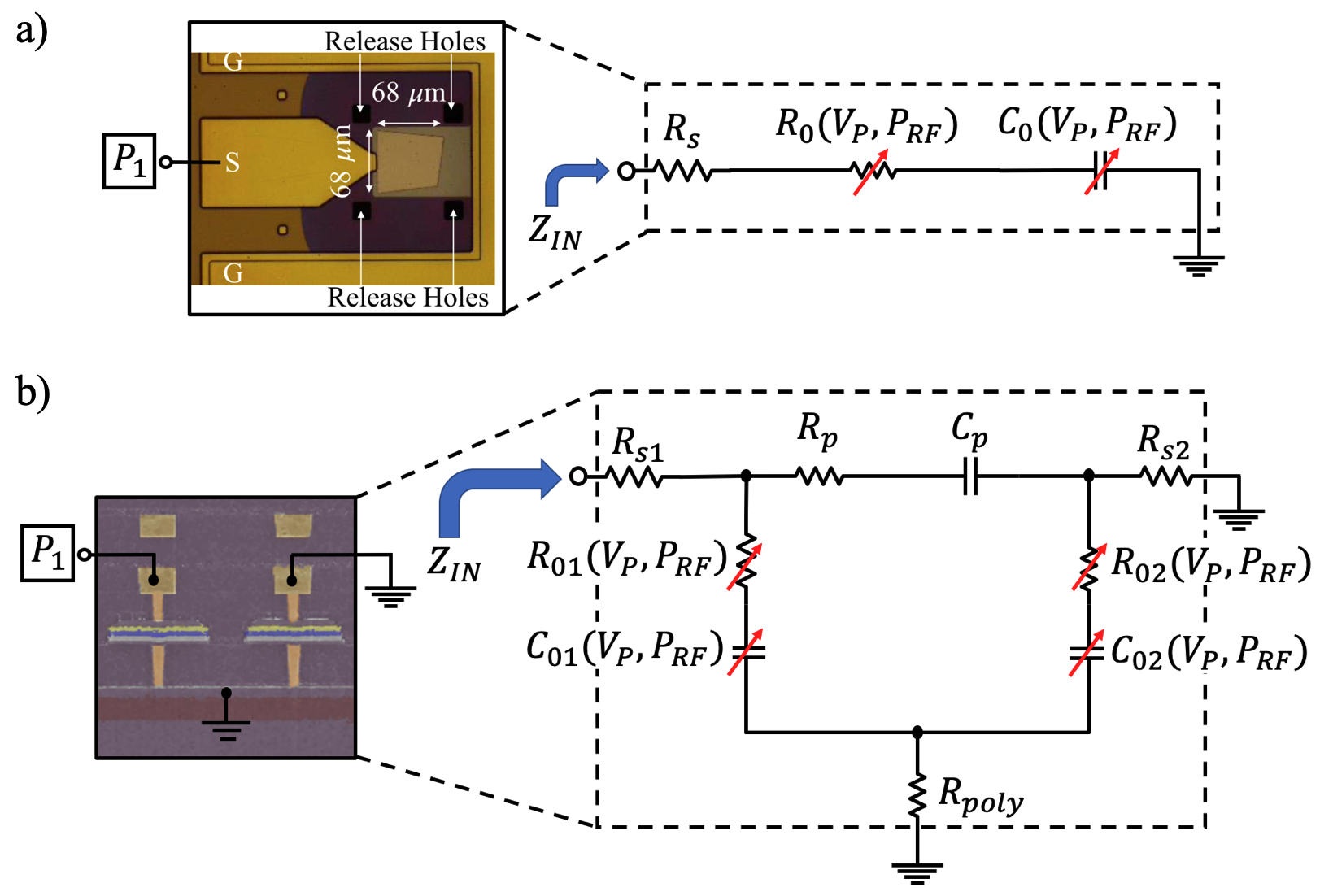}
%\vspace{-2mm}
\caption{Comparison of equivalent circuits for low frequency extraction of electrical parasitic elements in (a) BST FBAR and (b) FeCAP Resonator.}
\label{LowFreqC0extract}
%\vspace{-4mm}
%\vspace{-6mm}
\end{figure}

Tunable FBAR devices \cite{BST_FBAR} based on ferroelectric transducers such as Ba$_{x}$Sr$_{1-x}$TiO$_{3}$(BST) in its paraelectric phase have previously been modeled using both the mBVD \cite{BST_FBAR_MBVD1}\cite{BST_FBAR_MBVD2} and 1-D acoustic transmission-line models \cite{BST_FBAR_Masons1}\cite{BST_FBAR_Masons2}. MBVD models have also been used in the context of released PZT resonators \cite{PZT_released_MBVD1}\cite{PZT_released_MBVD2} but no attempt has yet been made to match the model consisting of bias-dependent parameters to the measured response characteristics under different operating conditions. The mBVD model extraction of ferroelectric BST FBARs relies on low-frequency extraction of the transducer capacitance $C_{0}$, which can be obtained from the imaginary part of input impedance as shown in Fig.\ref{LowFreqC0extract}(a). The extracted capacitance is then fitted to the well-known closed-form expression developed for BST varactors \cite{BST_Varactor}. Unlike the BST FBAR, the primary impediment to the direct extraction of transducer capacitances $C_{01}$ and $C_{02}$ in unreleased CMOS ferroelectric resonators is the presence of feedthrough paths between the input and output ports as shown in Fig.\ref{LowFreqC0extract}. Due to this, the imaginary part of the input impedance at low frequencies does not yield the correct values of the transducer capacitances. An additional challenge is that, unlike in paraelectric-phase BST, capacitance of ferroelectric Pb$_{x}$Zr$_{1-x}$TiO$_{3}$(PZT) exhibits a hysteretic nature which requires separate modeling treatment \cite{HysteresisModeling} for compatibility with RF simulation tools. The remainder of the extraction procedure \cite{TuneableFBARs} for the motional branch components ($R_{m}$, $L_{m}$ and $C_{m}$) is based on the modeling of the series and parallel resonance frequencies $f_{s}$ and $f_{p}$ with changing bias and extraction of the motional, unloaded Q $Q_{UL}$. In the case of low-$k_{eff}^2$ and moderate feedthrough such as in this work, it is not possible to uniquely determine the series resonant frequency $f_{s}$ for certain bias conditions since the resonance peak height is smaller than 3 dB. A different modeling and extraction procedure is needed for unreleased CMOS-MEMS ferroelectric resonators. Previous attempts at modeling the small-signal behaviour of these devices \cite{he_switchable_2019,he_tunable_2020} have been restricted to a limited number of biasing conditions and do not capture the full experimentally observed behaviour.

Mason's model implementations, such as those in \cite{BST_FBAR_Masons1}, are more amenable to FBAR-like devices consisting of different materials in the device stack. Similarly, other comprehensive models such as that in \cite{BST_FBAR_Masons2} make use of elastic and electrostrictive equivalent model slices based on full-field formulations and incorporation of nonlinear electrostrictive coupling terms. However, this approach is not applicable in the case of unreleased BEOL ferroelectric CMOS-MEMS devices due to more complicated geometries and mode shapes with a multitude of materials. As such, an augmented mBVD behavioural model as depicted in Fig.\ref{Fig2}(a) has been chosen to accurately capture the bias-dependent small-signal performance of the devices under consideration. Additionally, augmentation of the mBVD model to adequately incorporate large-signal nonlinearities has been studied previously in the context of FBAR devices further cementing the suitability of this approach for the same operation scenarios.

\subsection{Extraction Methodology}\label{supp_2}

As mentioned previously, the mBVD model in \cite{he_switchable_2019} cannot be used to capture performance variation with change in operating conditions due to absence of effects such as loss mechanisms in the ferroelectric material and capacitive coupling between the adjacent IDTs. These shortcomings result in a mismatch of greater than 2 dB between the measured and modeled transmission $S_{21}$ for different biases as well as RF power levels. The augmented mBVD model in Fig.\ref{Fig2}(a) significantly improves the discrepancy in the model's transmission response with respect to the measured de-embedded device data to within 1 dB for a majority of operating conditions through the incorporation of FeCAP transducer loss elements $R_{01}$ and $R_{02}$ in series with the corresponding intrinsic transducer capacitances $C_{01}$ and $C_{02}$. These loss elements, just like the capacitances themselves, are $V_{P}$ and $P_{RF}$ dependent. The series $RC$ combination of $R_{P}$ and $C_{P}$ captures the inter-finger capacitance and associated dielectric loss of the BEOL dielectric and is invariant with $V_{P}$ and $P_{RF}$ applied to the device.
The resistances $R_{s1}$ and $R_{s2}$ are used to model the routing resistance from the de-embedding plane to the device under test. The motional branch components of the model including motional resistance $R_{m}$, motional inductance $L_{m}$, and motional capacitance $C_{m}$ are interrelated through mechanical resonance frequency $f_{0}$ (angular frequency $\omega_{0}$) and mechanical Q $Q_{UL}$ as follows (the bias and RF power dependence of each of these quantities has been omitted for simplicity):

%\begin{subequations} \label{eq:motionalBranchEqs}
%    \begin{equation}
%        C_{m} = \frac{1}{\omega_{0}R_{m}Q_{UL}}
%    \end{equation}
%    \begin{equation}\vspace{2mm}
%       L_{m} = \frac{1}{\omega_{0}^2C_{m}}
%    \end{equation}
%\end{subequations}

\noindent\begin{minipage}{.5\linewidth} \vspace{4mm}
\begin{equation} \label{eq:Cm1}
    C_{m} = \frac{1}{\omega_{0}R_{m}Q_{UL}}
\end{equation}
\end{minipage}%
\begin{minipage}{.5\linewidth} \vspace{4mm}
\begin{equation} \label{eq:Lm1}
    L_{m} = \frac{1}{\omega_{0}^2C_{m}}
\end{equation}
\end{minipage} \vspace{2mm}

The magnitude of piezoelectric coupling coefficients $\eta_{1}$ and $\eta_{2}$, corresponding to the ferroelectric transducers, are implicitly included in the motional components as follows:

\noindent\begin{minipage}{.33\linewidth}
\begin{equation} \label{eq:Cm}
    C_{m} = \frac{\eta_{1}\eta_{2}}{C_{m}'}
\end{equation}
\end{minipage}%
\begin{minipage}{.33\linewidth}
\begin{equation} \label{eq:Lm}
    L_{m} = \frac{L_{m}'}{\eta_{1}\eta_{2}}
\end{equation}
\end{minipage}
\begin{minipage}{.33\linewidth}
\begin{equation} \label{eq:Rm}
    R_{m} = \frac{R_{m}'}{\eta_{1}\eta_{2}}
\end{equation}
\end{minipage} \vspace{2mm}

where $R_{m}'$, $L_{m}'$ and $C_{m}'$ are the motional branch circuit parameters independent of electromechanical coupling. This explicit separation of motional and electromechanical transducer parameters is done to allow direct extraction of $R_{m}$ from measurements across bias and RF powers and insertion directly into the model as initial values prior to optimization.

\begin{figure}[t!]
\centering
\includegraphics[width=0.5\textwidth]{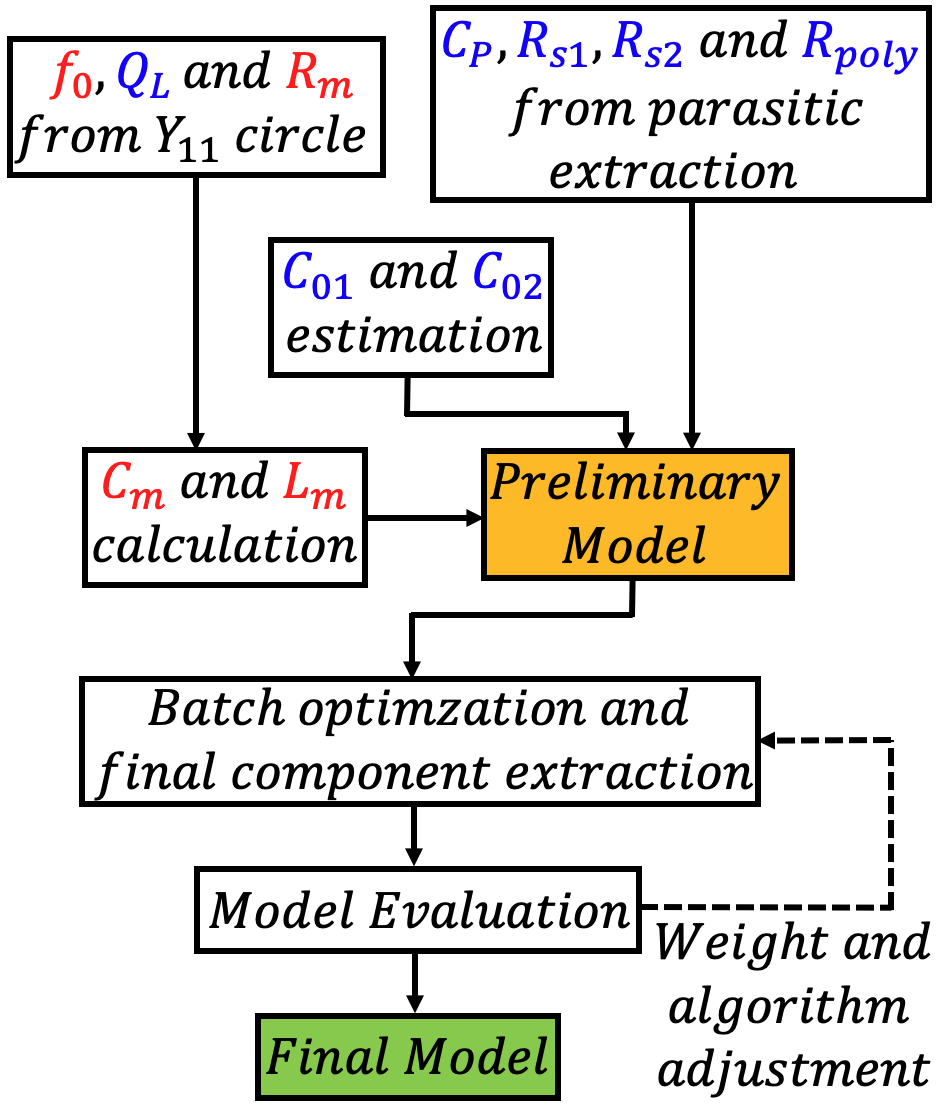}
%\vspace{-2mm}
\caption{Extraction flow for the resonator based on de-embedded measurement data.}
\label{SSigExtFlow}
%\vspace{-4mm}
%\vspace{-6mm}
\end{figure}

Once the model structure has been finalized and all the necessary circuit components with a physical basis have been included, a parameter/component extraction is required to build a preliminary model for the resonator. The first step in this procedure (as seen in Fig.\ref{SSigExtFlow}) is the estimation of the motional branch parameters from the measured data across biases after open and short de-embedding. To overcome the modeling challenges mentioned in the previous section, an admittance circle based methodology is adopted \cite{JEYLee_feedthrough}. If only the motional branch of the model is considered then the admittance looking into the branch is given by:

\begin{equation}\label{eq:Y11prime}
    Y_{11}'=G(\omega)+jB(\omega)=\frac{1}{R_{m}+j\omega L_{m}+1/j\omega C_{m}}
\end{equation}

where $G$ and $B$ are the conductance and susceptance respectively which are given by:
\begin{subequations}
\begin{equation}\label{eq:G}
    G(\omega)=\frac{\omega^{2}C_{m}^{2}R_{m}}{(1-\omega^{2}L_{m}C_{m})^{2}+\omega^{2}C_{m}^{2}R_{m}^{2}}
\end{equation}
\begin{equation}\label{eq:B}
    B(\omega)=\frac{(1-\omega^{2}L_{m}C_{m})\omega C_{m}}{(1-\omega^{2}L_{m}C_{m})^{2}+\omega^{2}C_{m}^{2}R_{m}^{2}}
\end{equation}
\end{subequations}

from these equations the admittance circular locus equation can be obtained as follows:

\begin{equation}\label{eq:circle1}
    \left( G(\omega)-\frac{1}{2R_{m}} \right)^{2}+B^{2}(\omega)=\left( \frac{1}{2R_{m}} \right)^{2}
\end{equation}

\begin{figure}[t!]
\centering
\includegraphics[width=0.62\textwidth]{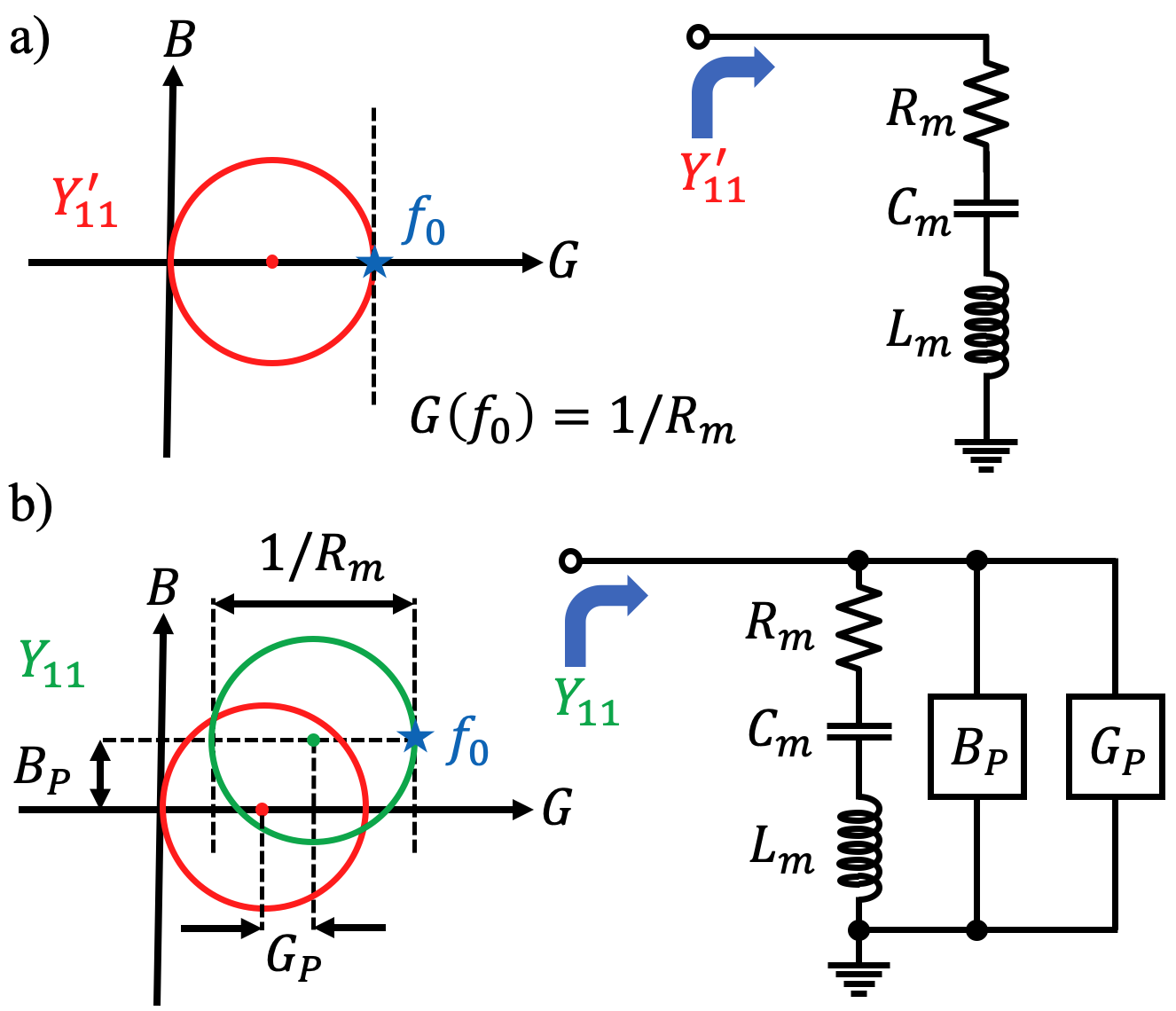}
%\vspace{-2mm}
\caption{Schematic depicting the $Y_{11}$ admittance circle methodology for extracting equivalent circuit parameters from RF measured data.}
\label{Y11circles}
%\vspace{-4mm}
%\vspace{-6mm}
\end{figure}

This admittance circle has been depicted in the Nyquist plot schematic of Fig.\ref{Y11circles}(a). The frequency $f_{0}$ corresponds to the mechanical resonance frequency of the mode under consideration ($B=0$). The inverse of the circle's diameter is equivalent to $R_{m}$. Parallel feedthrough branches corresponding to IDT coupling capacitance $C_{P}$ and the intrinsic ferroelectric capacitances $C_{01}$, $C_{02}$ as well as $R_{poly}$ together contribute an equivalent conductance $G_{P}$ and susceptance $B_{P}$ as can be seen in Fig\ref{Y11circles}(b). Since the resonance is high-Q and the feedthrough level is moderate, it can be assumed that $G_{P}$ and $B_{P}$ are invariant with frequency in the frequency range under consideration. Under this assumption, the circle is shifted in the Nyquist plane by $G_{P}$ and $B_{P}$ along the $G$ and $B$ axes, respectively. It should be noted that the diameter of the admittance circle ($Y_{11}$) does not change when additional components are added in parallel to the motional branch which is instrumental in the direct extraction of $R_{m}$ from measurement.

Furthermore, the point corresponding to the maximum conductance $G(\omega)$ is used to obtain the mechanical resonance frequency $f_{0}$. The maxima and minima points of the susceptance $B(\omega)$ correspond to the two -3dB frequencies. Using the obtained $f_{0}$ and the two -3dB frequencies, the loaded Q $Q_{L}$ can be obtained from the 3dB bandwidth: $Q_{L}=f_{0}/(f_{-3dBu}-f_{-3dBl})$. Under conditions of moderate feedthrough the -3dB frequencies deviate slightly from the series and parallel resonance frequencies ($f_{S}$ and $f_{P}$)\cite{JEYLee_feedthrough}. Using the relations highlighted in equations \ref{eq:Cm1} and \ref{eq:Lm1} the corresponding initial estimates of $C_{m}$ and $L_{m}$ can be obtained. The series routing resistances $R_{s1}$ and $R_{s2}$ also influence these calculations but the obtained values still serve as a good starting point for the optimization engines subsequently used. 

Using layout extraction (as in \cite{Rawat11GHz}), parasitic electrical components including $C_{P}$, $R_{s1}$, $R_{s2}$ and $R_{poly}$ can be estimated and fed back into the preliminary model. The de-embedding plane used for obtaining the device measurement characteristics using on-chip open and short structures is considered while estimating these parasitic elements. Although the transducer capacitances extracted from the ferroelectric P-E loop measurement in \cite{he_switchable_2019} do not capture the accurate values because of challenges enumerated in Section \ref{supp_1}, they are still used in the pre-optimization model to get a good initial estimate. Once the preliminary model has been prepared, a batch optimization is carried out using Keysight\textsuperscript{\textregistered} ADS across different bias voltages ($V_{P}$) and RF powers $P_{RF}$. The error function is appropriately set to minimize deviation in the model magnitude as well as phase response with respect to the measured de-embedded device characteristics. The model fit quality is subsequently evaluated and requisite adjustments in the optimization weights and algorithms are made to obtain the final model which fits as close as possible to the measured data at the bias points considered.   

\subsection{Model Benchmarking}\label{supp_3}

\begin{figure}[ht!]
  \centering
  \begin{tabular}{cc}
    \includegraphics[width=0.45\textwidth]{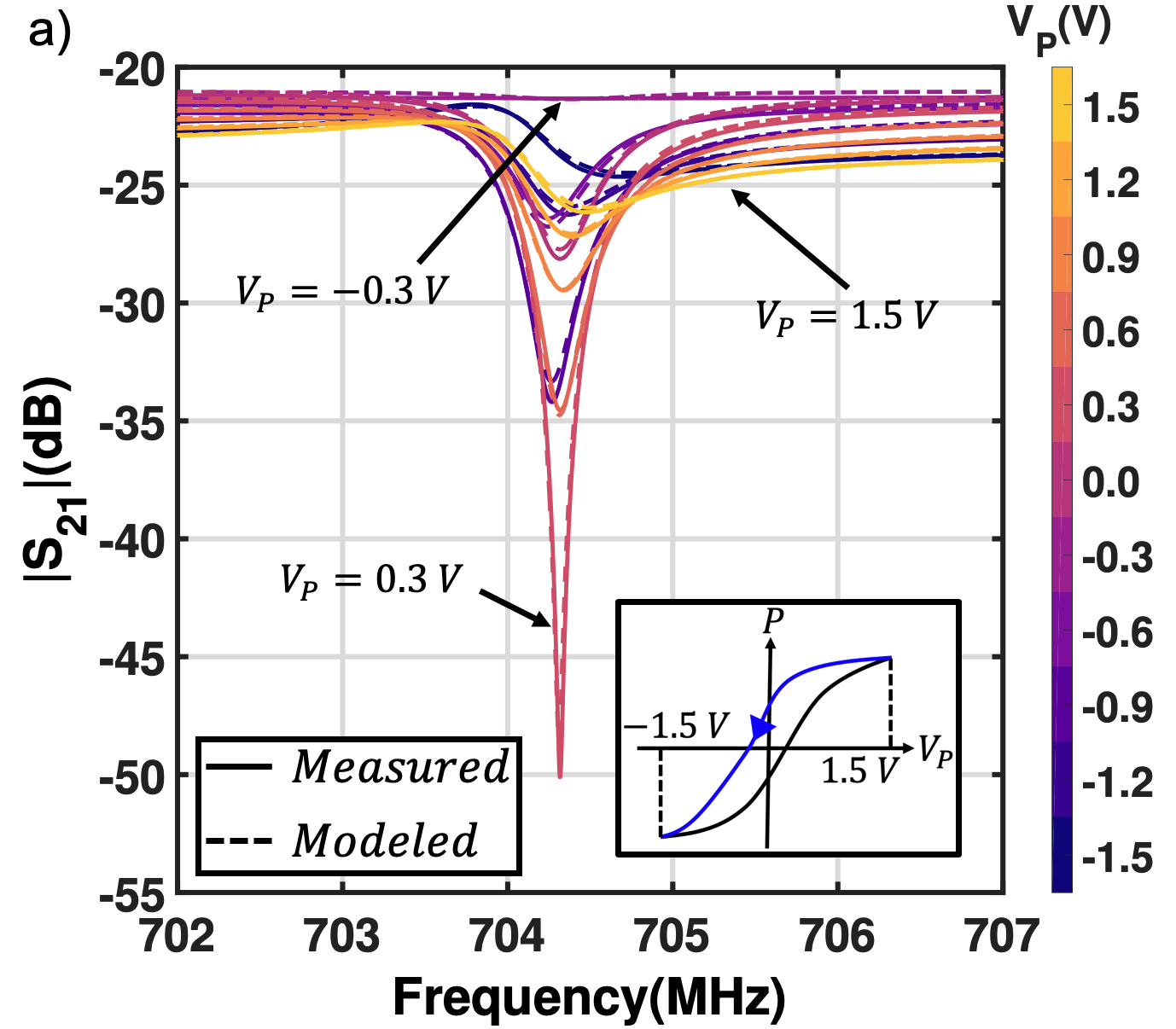} &
    \includegraphics[width=0.45\textwidth]{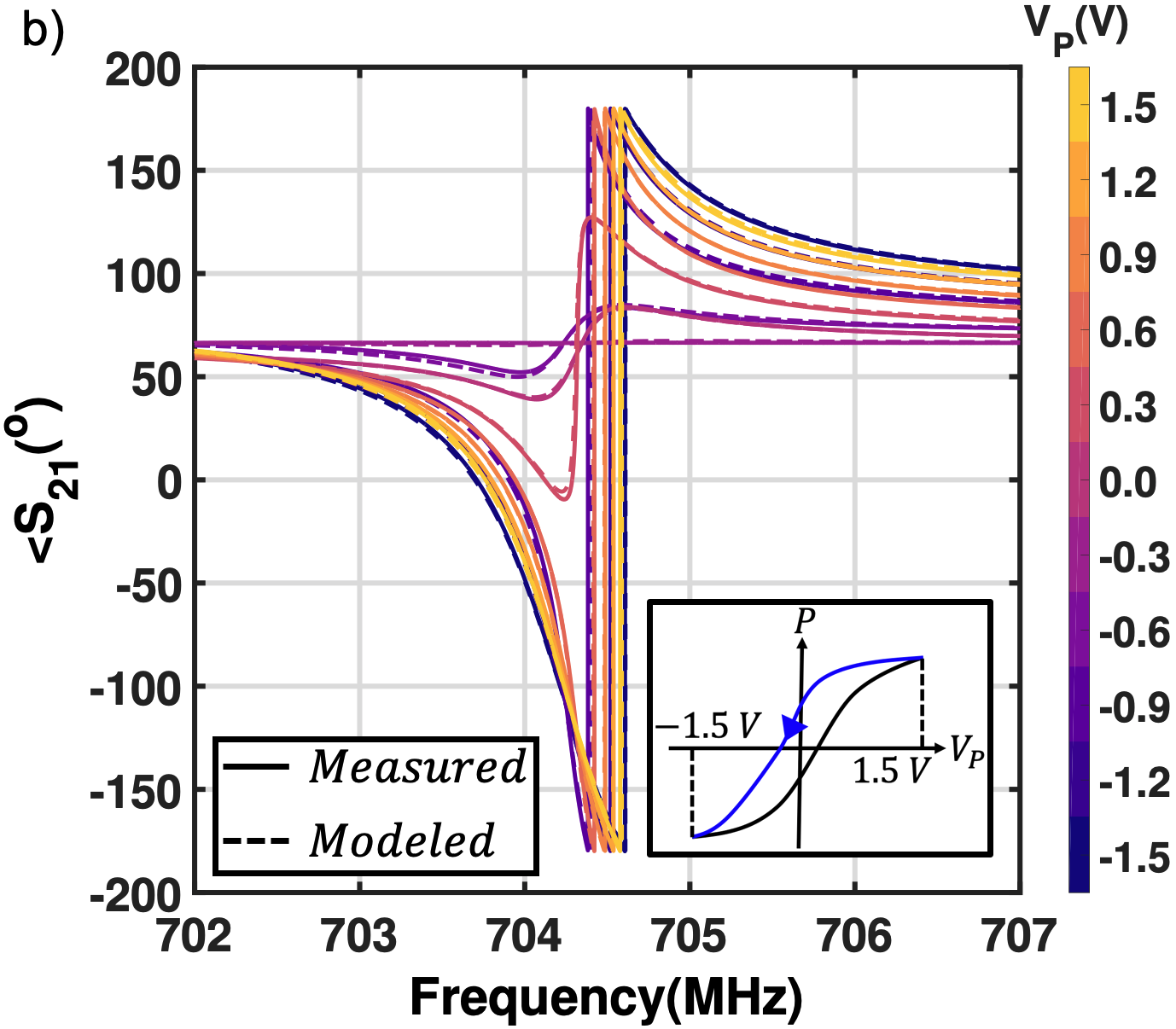} \\
    \includegraphics[width=0.45\textwidth]{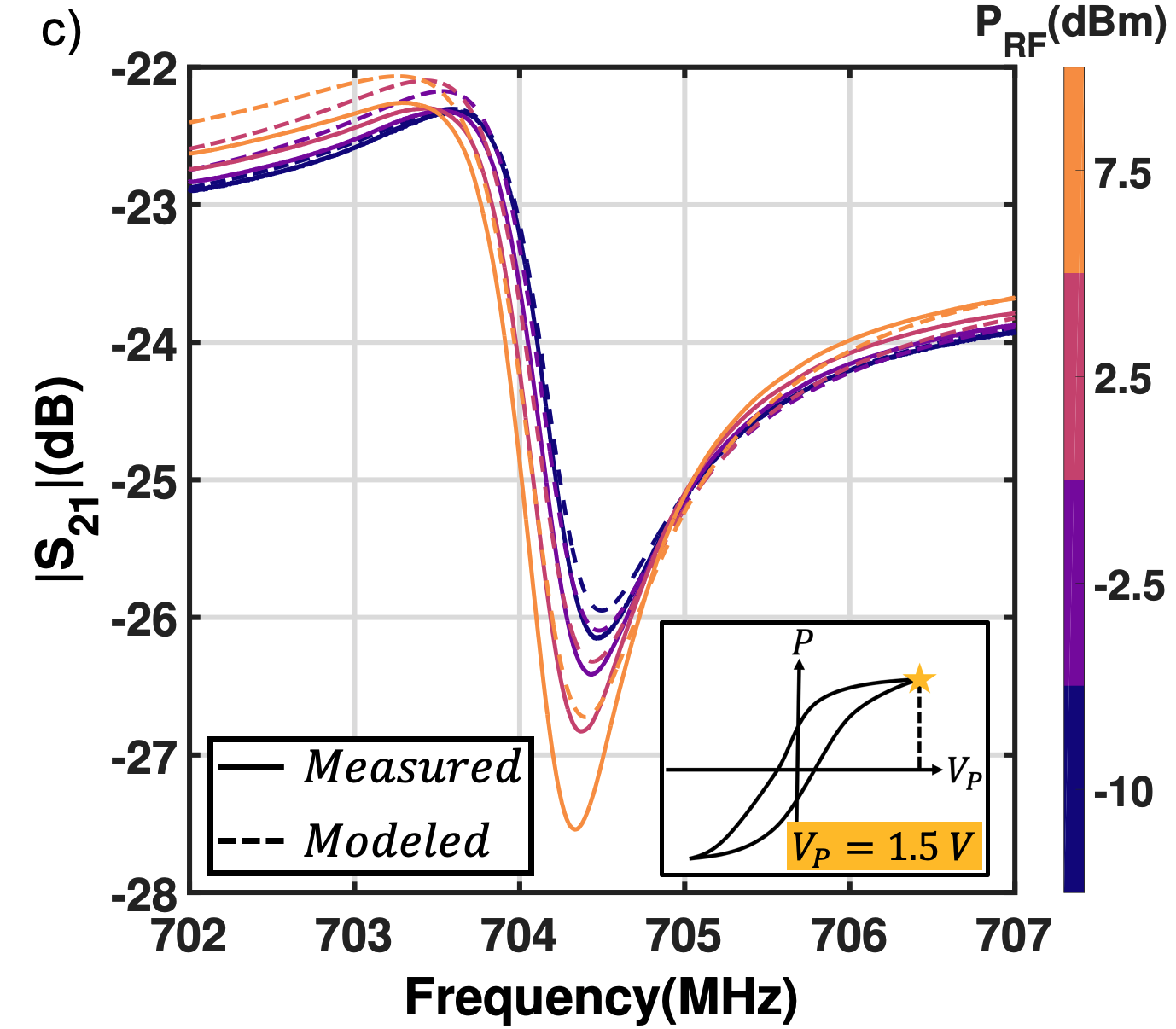} &
    \includegraphics[width=0.45\textwidth]{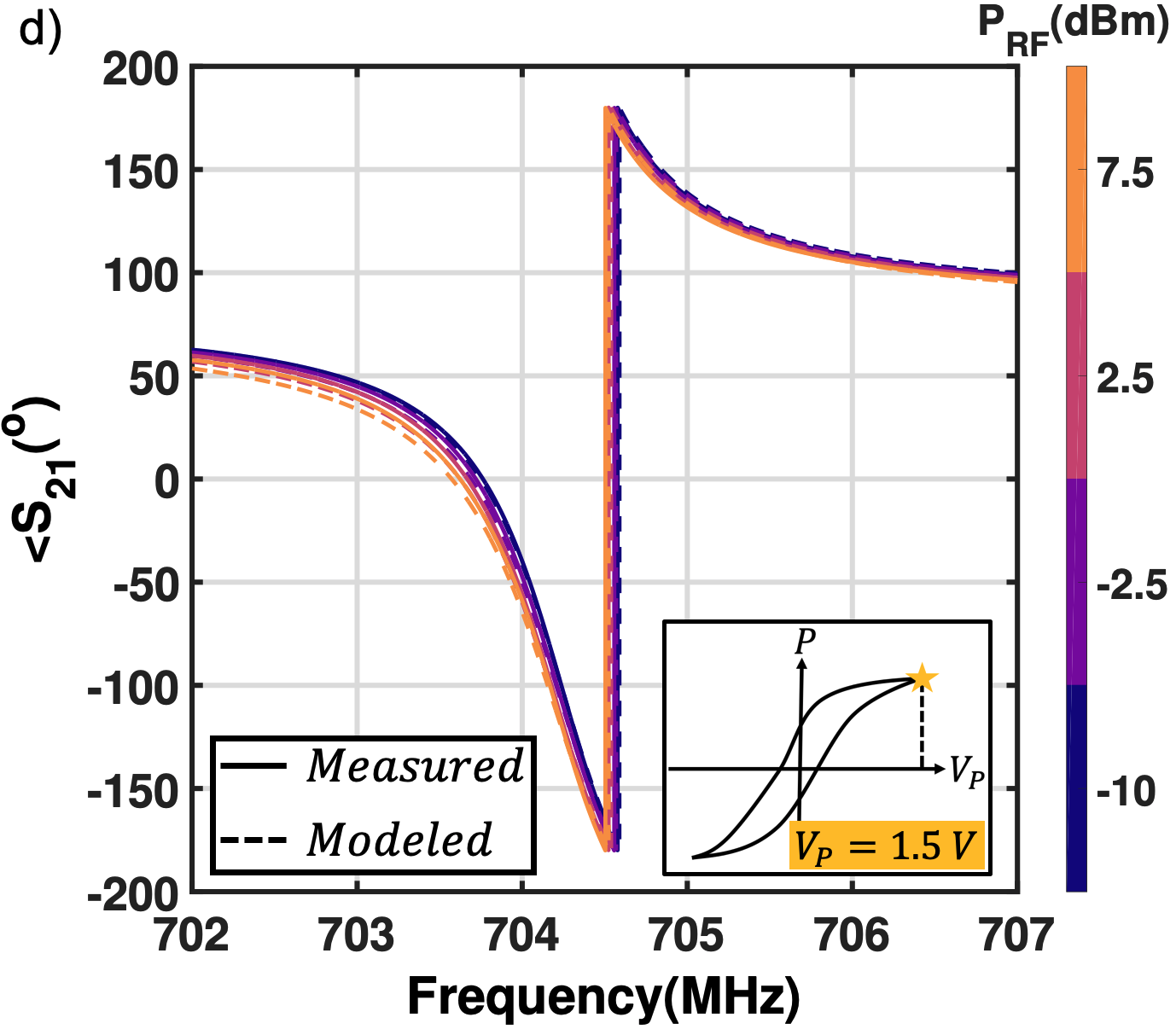} \\
  \end{tabular}
  \caption{Comparison of measured and modeled magnitude and phase characteristics for (a), (b) small signal $P_{RF}$= -10 dBm and (c), (d) large signal $V_{P}$ = -0.3 V operation.}\label{supp_ModelBench}
\end{figure}

The comparison of the measured transmission magnitude and phase characteristics with respect to the model is shown in Fig.\ref{supp_ModelBench}(a) and (b) across different $V_{P}$ values traversing along the P-E loop. It can be observed that model matches closely with the measured characteristics. This demonstrates the efficacy of the small-signal model extraction methodology that has been described in the previous section. The extraction methodology remains unchanged for large signal operation since the relations governing the model operation remain constant with varying amplitude. This can be qualitatively understood in terms of a basic Volterra series formulation \cite{bookVolterra} for the large-signal transadmittance $Y_{21}$ response of the equivalent circuit model of Fig.\ref{Fig2}(a). Assuming that an input voltage of amplitude $A$ and frequency $\omega_{1}$ given by $v_{in}(t)=A\exp(j\omega_{1} t)$ is applied to port 1 (P1) of the model, then the output current at port 2 (P2) may be represented in the form:

\vspace{-2mm}
\begin{multline}\label{eq:Volterra}
    i_{out}(t)=H_{1}(\omega_{1})A\exp(j\omega_{1} t) + H_{2}(\omega_{1},\omega_{1})A^{2}\exp(2j\omega_{1} t) \\
    + H_{3}(\omega_{1},\omega_{1},\omega_{1})A^{3}\exp(3j\omega_{1} t) + ...
\end{multline}
%\vspace{1mm}

where the transfer functions $H_{n}$ represent the Fourier Transform of the corresponding $n$-th order Volterra kernels corresponding to the transadmittance $Y_{21}$. With an increasing input amplitude $A$, the overall $Y_{21}$ magnitude and phase response of the device deviates from the linear response due to $H_{1}(\omega_{1})$ under small-signal conditions due to the action of higher order kernels $H_{2}(\omega_{1},\omega_{1})$, $H_{3}(\omega_{1},\omega_{1},\omega_{1})$ and so on. Each remaining individual Volterra kernel in itself is dependent upon the nonlinear coefficients of the various components in the model as well as other kernels. The primary goal is to determine these nonlinear coefficients of the individual components using the extracted component values at different RF power levels to construct a nonlinear large-signal model of the resonator.

To make the nonlinear resonator model amenable to circuit simulation, it is desirable to make the nonlinear model components voltage- and current-dependent rather than $P_{RF}$ dependent. The technique highlighted in \cite{LSSP_BSTFBAR}, applied to a standard 1-port FBAR model, is augmented and utilized to capture the nonlinear behaviour of the motional as well as transducer components. The behavior of each of these components is captured using a power series as follows:

\begin{subequations}
\begin{equation}\label{eq:Rm_NL}
    R_m = R_{m,0}+R_{m,1}I_{m}+R_{m,2}I_{m}^2+R_{m,3}I_{m}^3+R_{m,4}I_{m}^4
\end{equation}
\begin{equation}\label{eq:Lm_NL}
    L_m = L_{m,0}+L_{m,1}I_{m}+L_{m,2}I_{m}^2+L_{m,3}I_{m}^3+L_{m,4}I_{m}^4
\end{equation}
\begin{equation}\label{eq:Cm_NL}
    C_m = C_{m,0}+C_{m,1}V_{Cm}+C_{m,2}V_{Cm}^2+C_{m,3}V_{Cm}^3+C_{m,4}V_{Cm}^4
\end{equation}
\end{subequations}

where $R_{m,x}$, $L_{m,x}$ and $C_{m,x}$ ($x=0$ to $4$) are polarization-dependent nonlinear coefficients, $I_{m}$ is the current through the motional branch and $V_{Cm}$ is the voltage across the motional capacitor. The frequency-dependent $I_{m}$ and $V_{Cm}$ are used to calculate the nonlinear coefficients for all the polarization voltages under consideration. The components are then implemented using equations based Symbolically Defined Device (SDD) components in Keysight\textsuperscript{\textregistered} ADS. Subsequently, harmonic balance (HB) large-signal S-parameter simulation is carried out to compare the model output against the measured device response. From Fig.\ref{supp_ModelBench}(c) and (d), it can be seen that for $V_{P}$ of 1.5V, the model matches closely with measured device S-parameters across multiple $P_{RF}$ values. Optimization of fitting to the power series across different $V_{P}$s ensures that the model S-parameter magnitude is within 1 dB of the measured response magnitude. Similarly, it is ensured that the model phase does not deviate beyond $5^{o}$ from the measured phase characteristics.

\bibliography{sn-bibliography}% common bib file


\end{document}